\documentclass[times,sort&compress,3p]{elsarticle}
\journal{Journal of Multivariate Analysis}
\usepackage[labelfont=bf]{caption}

\usepackage{amsmath,amsfonts,amssymb,amsthm,booktabs,color,epsfig,graphicx,hyperref,url}
\usepackage{enumitem}
\usepackage{array}

\theoremstyle{plain}

\newtheorem{propr}{Property}

\theoremstyle{definition}

\newtheorem{remark}{Remark}
\newtheorem{example}{Example}

\providecommand{\bo}{\mathbf}
\providecommand{\bs}{\boldsymbol}
\newcommand{\V}{\bs V}
\newcommand{\X}{\bs X}
\newcommand{\x}{\bs x}

\newcommand{\Z}{\bs Z}
\providecommand{\PP}{\mathcal{P}_p}
\providecommand{\SP}{\mathcal{SP}_p}
\providecommand{\cov}{\mathrm{COV}}
\providecommand{\covg}{\mathrm{GCOV}}
\providecommand{\mcd}{\mathrm{MCD}}
\providecommand{\mrcd}{\mathrm{MRCD}}
\providecommand{\diag}{\mathrm{diag}}
\providecommand{\rank}{\mathrm{rank}}
\providecommand{\Null}{\mathrm{null}}
\providecommand{\range}{\mathrm{range}}

\providecommand{\Proj}{\bo \mathrm{Proj}}
\providecommand{\argmax}{\mathrm{argmax}}

\DeclareMathSymbol{\Mu}{\mathalpha}{operators}{"4D}

\newcommand{\pkg}[1]{\texttt{#1}}
\newcommand{\proglang}[1]{\textsf{#1}}

\begin{document}

\begin{frontmatter}

\title{Generalized implementation of invariant coordinate selection  with positive semi-definite  scatter matrices}

\author[1]{Aurore Archimbaud}

\address[1]{TBS Business School, 1 Place Alphonse Jourdain, 31000 Toulouse, France}
\cortext[mycorrespondingauthor]{Corresponding author. Email address: \url{a.archimbaud@tbs-education.fr}}

\begin{abstract}
Invariant coordinate selection is an unsupervised multivariate data transformation useful in many contexts such as outlier detection or clustering. It is based on the simultaneous diagonalization of two affine equivariant and positive definite scatter matrices. Its classical implementation relies on a non-symmetric eigenvalue problem by diagonalizing one scatter relatively to the other. In case of collinearity, at least one of the scatter matrices is singular, making the problem unsolvable. To address this limitation, three approaches are proposed using: a Moore-Penrose pseudo inverse, a dimension reduction, and a generalized singular value decomposition. Their properties are investigated both theoretically and through various empirical applications. Overall, the extension based on the generalized singular value decomposition seems the most promising, even though it restricts the choice of scatter matrices to those that can be expressed as cross-products. In practice, some of the approaches also appear suitable in the context of data in high-dimension low-sample-size data. 
\end{abstract}

\begin{keyword} 
Dimension reduction \sep
Generalized Eigenvalue Problem \sep
High-dimension \sep
Pseudo-inverse \sep
Singular scatters \sep
Singular value decomposition 
\MSC[2020] Primary 62H99 \sep 
Secondary 62-08 \sep 
Tertiary 65F99 

\end{keyword}

\end{frontmatter}

\section{Introduction\label{sec:intro}}
Invariant Coordinate Selection (ICS) is a powerful unsupervised multivariate method designed to identify the structure of multivariate datasets on a subspace. It relies on 
the summary of the difference between two affine equivariant and positive definite scatter matrices $\V_1$ and $\V_2$, by examining the eigenvalues and eigenvectors of one matrix with respect to the other. This joint diagonalization is particularly relevant as a dimension reduction tool prior to clustering \cite{alfons_tandem_2024} or outlier detection \cite{archimbaud_ics_2018}. It goes beyond the well-known Principal Components Analysis (PCA) method by not maximizing the inertia but optimizing a generalized kurtosis. More precisely, some theoretical results \cite{tyler_invariant_2009} proved that under some elliptical mixture models, the subspace spanned by the first and/or last components carries the information regarding the multivariate structure and recovers the Fisher discriminant subspace, regardless of the choice of scatter matrices. 

Typically, assuming that $\V_1  \in \mathcal{P}_p$ and $\V_2 \in \mathcal{P}_p$, with $\mathcal{P}_p$ be the set of all symmetric positive definite matrices of order $p$ and following the normalization given by \cite{tyler_invariant_2009}, the goal of ICS is to find the non-trivial eigenvalues $\rho_i$ and the corresponding eigenvectors $\bo b_i$ of $\V_2$ relative to  $\V_1$ for $i\in \{1,\ldots,p\}$:
\begin{equation}\label{def:ICSvect0}
\V_2 \bo b_i = \rho_i \V_1 \bo b_i.
\end{equation}
This simultaneous diagonalization corresponds to a generalized eigenvalue problem (GEP), which might be not so familiar to most readers. However, equivalently,  $\rho_i$ and $\bo b_i$  are the eigenvalue and the corresponding eigenvector of $\V_1^{-1}\V_2$. So, the problem~(\ref{def:ICSvect0}) can be simplified to a more usual non-symmetric eigenvalue problem (EVP) by directly diagonalizing the non-symetric matrix $M = \V_1^{-1} \V_2$:
\begin{equation}\label{def:ICSvect}
\V_1^{-1} \V_2 \bo b_i = \rho_i  \bo b_i.
\end{equation}
As stated in \cite{tyler_invariant_2009}, without loss of generality, we can choose to use the following normalization:
\begin{itemize}[noitemsep,topsep=0pt,parsep=0pt,partopsep=0pt]
	\item $\bo b_i^\top \V_1\bo b_j=0$ for $i\neq j$ and $\bo b_j^\top \V_1\bo b_j=1$ for $i = j$, with $i, j \in \{1,\ldots,p\}$,
	\item  $\bo b_i^\top \V_2\bo b_j=0$ for $i\neq j$ and $\bo b_j^\top \V_2\bo b_j=\rho_j$ for $i =  j$, with $i, j \in \{1,\ldots,p\}$.
\end{itemize}

Equivalently, this problem can be re-written as:
\begin{equation}\label{def:ICSmatrix}
\bo B \V_1 \bo B^\top = \bo I_p \quad \mbox{and} \quad \bo B \V_2 \bo B^\top = \bo D,
\end{equation}
where $^\top$ denotes the transpose operation, $\bo D$ is a diagonal matrix with decreasing diagonal elements $\rho_1\geq \dots \geq \rho_p>0$, which correspond to the eigenvalues of $\V_1^{-1}\V_2$ and $\bo B$ contains the corresponding eigenvectors $\bo b_1, \ldots,\bo b_p$ as its rows.

Finally, as noted by \cite{tyler_invariant_2009}, the eigenvalues $\rho_i$, for $i \in \{1,\ldots,p\}$ and the eigenvectors $\bo b_1, \ldots,\bo b_p$ can also be sequentially defined by solving the successive maximization or minimization problems of the ratio: 
\begin{equation}\label{def:ICSratio}
\mathcal{K}(\bo b) = \frac{\bo b^\top  \V_2 \bo b}{\bo b^\top  \V_1 \bo b},
\end{equation}
where $\rho_1$ is the maximal possible value of $\mathcal{K}(\bo b) $ over $\bo b \in \mathcal{R}^p$ which is achieved in the direction of the eigenvector $\bo b_1$. Similarly, the minimal value of $\mathcal{K}(\bo b)$ is $\rho_p$, which is achieved in the direction of $\bo b_p$. More generally, as stated in the  equations (15) and (16) in \cite{tyler_invariant_2009}, we have:
	\[
		\sup \{\mathcal{K}(\bo b) |\bo b \in \Re^p, \bo b'  \V_1 \bo b_j, j=1, \dots, m-1\}=\rho_m
	\]
with the supremum being obtained at $\bo b_m$, and
	\[
		\inf \{\mathcal{K}(\bo b) |\bo b \in \Re^p, \bo b'  \V_1 \bo b_j, j=m+1, \dots, p\}=\rho_m
	\]
with the infimum being obtained at  $\bo b_m$.

As $\mathcal{K}(\bo b)^2$ can be viewed as a generalized measure of kurtosis, it follows that $\rho_i$ has a useful interpretation. Thus the so-called invariant coordinates or components can reveal interesting structures, mostly the subspace spanned by the first and/or last components. They are obtained as: $\Z_n =(\X_n - \bo 1_n\bs T(\X_n)^\top)\bs B({\X_n})^\top$, where $\X_n =(\x_1,\ldots,\x_n)^\top \in \mathcal{R}^{n \times p}$ is a $p$-variate sample with $n$ observations, $\bo 1_n$ denotes an $n$-variate vector full of ones, and $\bs T(\X_n)$ denotes a location estimator, usually the one that goes along with $\V_1$.

~

Nowadays, more and more data are easily collected, and collinearity issues arise more frequently even when the number of observations is still higher than the number of variables. In this context, the scatter estimates, such as the variance-covariance matrix, might become (exactly or nearly) singular. Performing ICS becomes very challenging as we might not be able to compute: (i) one or both scatter matrices and (ii)  the inverse of $\V_1$ as required to solve the GEP~(\ref{def:ICSvect}). So in practice, the requirement of positive definiteness for the two scatter matrices is a limiting issue in the application of ICS. To the best of my knowledge, only a few attempts have been made to circumvent this issue. In case of numerical instability due to nearly singular scatter matrix, \cite{archimbaud_numerical_2023} propose a new implementation of ICS based on a pivoted QR factorization restricted to the case of one-step M-scatter matrices. In case of high dimension low sample size data (HDLSS), they also suggest leveraging row and column pivoting to define a more stable rank-revealing procedure of the data that allows the computation of ICS when at least one of the scatter matrices is singular. This idea of preprocessing the data through dimension reduction to reduce its rank, has also been used by \cite{fischer_subgroup_2017} but no theoretical investigation has been made so far.

More generally, collinearity is a long-standing issue in multivariate analysis since quite a lot of methods rely on the simultaneous diagonalization of two or more scatter matrices \cite{nordhausen_usage_2022}, for which the non-singularity of a scatter matrix is required.  In the past, a simple idea was to perform a selection of variables if many variables are known to be non-relevant. However, this procedure can induce the deletion of a substantial number of variables to obtain a convenient number of dimensions on which a scatter estimator can be defined, leading to a potential loss of information. One of the most well-known methods that solves a GEP of two scatter matrices is the classical Fisher Linear Discriminant Analysis (LDA), which maximizes the separation ratio of the between and the within-group covariance matrices $\bs \Sigma_B$ and $\bo  \Sigma_W$. In case of collinearity, the within-group covariance matrix  $\bo  \Sigma_W$ could become singular. So, the maximization problem cannot be performed by solving the  $\bo  \Sigma_W^{-1} \bs \Sigma_B$ eigenvalue problem anymore. To overcome that issue, \cite{howland_generalizing_2004} and \cite{tebbens_improving_2007} review some solutions. Among others, the proposed approaches exploit the Moore-Penrose pseudo-inverse, the dimension reduction, or the Generalized Singular Value Decomposition (GSVD).
The objective of this paper is to adapt some of those approaches to generalizing ICS to the singularity issue caused by collinearity and to investigate theoretically and practically their properties and to provide \proglang{R} \citep{r_core_team_r_2023}  implementations as well.

The structure of this paper is as follows. In Section~\ref{sec:sdp}, we introduce a definition of ICS with semi-definite positive scatter matrices and we present the challenges associated with such a context. In Section~\ref{sec:implementation}, we propose three approaches to adapting ICS to the case of semi-definite positive scatter estimates based on the Moore-Penrose pseudo inverse (GINV), the dimension reduction (DR), and the generalized singular value decomposition (GSVD). We also investigate theoretically their properties in terms of: (i)~the criterion to optimize, (ii)~the affine invariance of the scores, and (iii)~the symmetry of the roles of the two scatter estimates $\V_1$ and $\V_2$. Section~\ref{sec:applications} focuses on different empirical applications to confirm or infirm the theoretical aspects. Finally, Section~\ref{sec:conclu} concludes the paper and discusses further perspectives.

\section{ICS with semi-definite positive scatter matrices\label{sec:sdp}}
In Subsection~\ref{subsec:scatters}, some scatter matrices are detailed and a more general definition is given in case it is semi-definite positive. Subsection~\ref{subsec:ICS} generalizes ICS to positive semi-definite scatter matrices,  while Subsection~\ref {subsec:challenges} presents the main challenges associated with it.

\subsection{Scatter matrices}\label{subsec:scatters}
Let $\mathcal{P}_p$ be the set of all symmetric positive definite matrices of order $p$, $\mathcal{SP}_p$ be the set of all symmetric positive semi-definite matrices of order $p$ and $\bo X_n \in \mathcal{R}^{n \times p}$, the initial data containing $n$ observations, characterized by $p$ variables. Generally, a scatter matrix is defined as a $p \times p$ scatter matrix $\V(\X_n) \in \mathcal{P}_p$ which is affine equivariant in the sense that: 
$$
\V(\X_n \bo A + \bo 1_n \bo \gamma^\top) = \bo A^\top \V(\X_n) \bo A,
$$

where $\bo A$ is a full rank $p \times p$ matrix, $\bo \gamma$ a $p$-vector and $\bo 1_n$ an $n$-vector full of ones. Among the most common ones there are the regular covariance matrix:
$$
\cov(\X_n)= \frac{1}{n-1} \sum_{i=1}^{n} (\x_i-\bar{\x})(\x_i-\bar {\x})^\top,
$$

where $\bar {\x}$ denotes the empirical mean, and the so-called scatter matrix of fourth moments:
$$
\cov_4(\X_n) = \frac{1}{(p+2)n} \sum_{i=1}^{n} r_i^2 (\x_i-\bar {\x})(\x_i-\bar {\x})^\top,
$$
where $r_i^2 = (\bo x_i - \bar {\bo x})^\top\cov(\X_n)^{-1}(\bo x_i - \bar {\bo x})$ is the classical squared Mahalanobis distance. It is well known that those two scatter matrices are not robust against non-normality and the presence of outliers. One of the most widely-used robust alternatives is the minimum covariance determinant estimator ($\mcd$) \citep{rousseeuw_multivariate_1985}. For a tuning parameter $\alpha \in [0.5, 1]$, the $\mcd$ selects out of the $n$ observations those $n_\alpha = \lceil \alpha n \rceil$ observations $\x_{i_1}, \ldots, \x_{i_{n_\alpha}}$ for which the sample covariance matrix has the smallest determinant: 
$$
\mcd_{\alpha}(\X_n) = c_\alpha \frac{1}{n_\alpha} \sum_{j=1}^{n_\alpha} (\x_{i_j}-\bar{\x}_{\alpha,n})(\x_{i_j}-\bar{\x}_{\alpha,n})^\top, 
$$ 
where $\bar{\x}_{\alpha,n}$ is the sample mean of the selected set of observations and $c_\alpha$ is a consistency factor. To increase efficiency, it is often combined with a reweighting step, see for example \citet{hubert_minimum_2018} for more details.

~

Theoretically, we can consider scatter matrices $\V(\X_n)$ which are only symmetric positive semi-definite. If $\X_n$ is not of full rank, then $\cov(\X_n) \in \mathcal{SP}_p$. For robust scatter matrices based on a subset of observations lying on a subspace of lower dimension than the entire space, $\V(\X_n)$ belongs to $\mathcal{SP}_p$. This arises for example in the presence of outliers in the orthogonal complement subspace (OC outliers). In this context, we can extend the definition of a scatter matrix $\V(\X_n) \in \mathcal{SP}_p$ which is said to be affine equivariant in the sense that:
$$
\V(\X_n \bo A + \bo 1_n \bo \gamma^\top) = \bo A^\top \V(\X_n) \bo A,
$$
for any $\bo A$ and $\bo \gamma$ as previously defined. \citet[Lemma 2]{tyler_note_2010} proves that if all the data lie in some $r \leq p$-dimensional hyperplane, then affine equivariant location and scatter statistics are essentially affine equivariant statistics defined on this hyperplane. For  a $p \times (p-r)$ matrix $\bo M$ of rank $p-r$ and $\bo m \in \mathcal{R}^{p-r}$, let us defined the hyperplane $\mathcal{H}(M,m)=\{\x \in  \mathcal{R}^{p} ~|~\bo M^\top\x=\bo m\}$, then for any affine equivariant scatter matrix $\V(\X_n)$: $
\bo M^\top \V(\X_n) \bo M = \bo 0_{(p-r) \times (p-r)}
$, where $\bo 0_{j \times k}$ denote the $j \times k$ matrix of all zeroes. In addition the lemma states that if $\bo L$ is any $p \times r$ matrix such that $\bo A =  \begin{pmatrix}\bo L & \bo M\end{pmatrix}$ is non singular then:
\begin{equation}\label{eqt:scatter}
\V(\X_n \bo A ) = \begin{pmatrix}
 \V^{(r)}(\X_n \bo L) & \bo 0_{r \times (p-r)} \\
\bo 0_{(p-r) \times r} & \bo 0_{(p-r) \times (p-r)}
\end{pmatrix},
\end{equation}
where  $\V^{(r)}(\bo Y)$ is a scatter matrix for $n \times r$ data matrices $\bo Y$.  \citet{tyler_note_2010} also note that if the data is in general position\footnote {Data is in general position if there is no subset of $k$ observations lying on a subspace of dimension $k-2$, with $k\leq p+1$ and $p$ denotes the number of variables.}, then all  affine equivariant scatter statistics, symmetric in the observations\footnote{i.e. $\V(\bo Q \X_n )=\V(\X_n)$ for any permutation matrix $\bo Q$ of order $n$ and $\X_n \in \mathcal{R}^{n \times p}$, the initial data containing $n$ observations, characterized by $p$ variables.}, are proportional to the variance-covariance matrix. In practice, if the data is multicollinear but with $n > p$ then the data is not in general position. If $n \leq p$, the situation depends on the data themselves but there are examples, mainly in the automotive field \citep[Chapter~5]{archimbaud_methodes_2018}, where the data is not in general position.

Another challenge is that many of the well-known robust affine equivariant scatter statistics, such as the M-estimators \citep{maronna_robust_1976} or the $\mcd$ are not well-defined when the data contains collinear variables. Usually, we can only define the variance-covariance matrix or the projection-based estimators \citep{donoho_breakdown_1992, maronna_bias-robust_1992, tyler_finite_1994} such as the Stahel-Donoho estimator \citep{stahel_breakdown_1981}. To overcome this issue, regularized estimators of scatter matrices have been proposed. For example, \cite{bickel2018projection} replace the within-group covariance matrix  $\bo  \Sigma_W$ with the regularized version  $\bo  \Sigma_W+\lambda \Omega$ where $\Omega$ is a penalty matrix and $\lambda$ a constant and \cite{loperfido2023kurtosis} suggests something similar for the covariance matrix. Among others, \cite{ollila_regularized_2014} present a regularization for the M-estimators and \cite{boudt_minimum_2020} for the $\mcd$, called minimum regularized covariance determinant estimator ($\mrcd$).

The scatter matrix based on the fourth moments $\cov4$ can also be defined in a more general context if the inverse of $\cov$ is replaced by its pseudo-inverse $\cov^+$:
$$
\covg_4(\X_n) = \frac{1}{(p+2)n} \sum_{i=1}^{n} r_i^2 (\x_i-\bar {\x})(\x_i-\bar {\x})^\top,
$$
where $r_i^2 = (\bo x_i - \bar {\bo x})^\top\cov^{+}(\X_n)(\bo x_i - \bar {\bo x})$ is the classical squared Mahalanobis distance. All those $p\times p$ scatter matrices belong to $\mathcal{SP}_p$ if $\X_n$ is not full rank and are no longer necessary affine equivariant. This is a well-known consequence and it is common to relax the affine invariance of the multivariate methods in case of singularity. For example, \cite{serfling_computationally_2013} propose to focus on some ``weak invariance" based on the relative ranking of the outlierness measures of the observations instead of requiring the same score value after an affine transformation.

For convenience, for the rest of the paper, the dependence on $\X_n$ is dropped from the different scatter matrices $\V(\X_n)$ when the context is obvious.

\subsection{ICS as a generalized eigenvalue problem}\label{subsec:ICS}
With the common definition of the ICS method, the two scatter matrices $\V_1$ and $\V_2$ should be definite positive to find a finite and nonzero eigenvalue $\rho_i$ to the eigenproblem~(\ref{def:ICSvect}): $\V_2 \bo b_i = \rho_i \V_1 \bo b_i \Leftrightarrow \V_1^{-1} \V_2 \bo b_i = \rho_i  \bo b_i$. The positive definiteness of $\V_1$ is required to compute its inverse whereas the positive definiteness of $\V_2$ ensures nonzero eigenvalues. 
If  $\V_1$ is singular and not proportional to $\V_2$, then the equivalence is not true anymore since $\V_1$ is not invertible. The problem can not be simplified to a non-symmetric eigenvalue problem (EVP) anymore and  so, we have to solve the initial Generalized Eigenvalue Problem  (GEP): 
\begin{equation}\label{GEP}
\V_2 \bo b_i = \rho_i \V_1 \bo b_i  \qquad \text{for~} i \in \{1,\dots,p\},
\end{equation} 
with $\V_1 \in  \mathcal{SP}_p$ and $\V_2 \in  \mathcal{SP}_p$ which are not necessarily of full ranks. If $\V_1$ and/or $\V_2$ is singular, that means that the null spaces of the two scatter matrices are not empty and they do not necessarily span the same subspace. Concretely, in this context, solving  the GEP of $\V_1$ and  $\V_2$, leads to consider the following cases:
\begin{itemize}[noitemsep,topsep=0pt,parsep=0pt,partopsep=0pt]
	\item if $\bo b_i \in \range(\V_1) \cap  \range(\V_2)$ then $\rho_i \in \mathcal{R}^{+*}$,
	\item if $\bo b_i \in \Null(\V_2) -   \Null(\V_1)$ then  $\rho_i=0$,
	\item if $\bo b_i \in \Null(\V_1) -   \Null(\V_2)$ then  $\rho_i=\infty$,
	\item	if $\bo b_i \in \Null(\V_1) \cap  \Null(\V_2)$ then any $\rho_i \in \mathcal{R}$ is a solution of the GEP. The corresponding eigenvectors are not well-defined and might cause stability issues for some algorithms. However, the structure of the data is not associated with those directions and so, we do not need to consider them further.
\end{itemize}
So, contrary to the classical ICS,  the directions $\bo b_i$ associated with infinite or zero eigenvalues should also be analyzed as they might highlight some of the structure of the data. Let us illustrate the challenges of considering semi-definite scatter matrices for ICS on one artificial data.

\subsection{Challenges with singular scatter matrices: an illustrative example}\label{subsec:challenges}
Let $\bo X=(X_1,\ldots,X_p)^\top$ be a $p$-multivariate real random vector and assume the distribution of  $\bo X$ is a mixture of two Gaussian distributions with different covariance matrices:

\begin{equation*}
{\bo X} \sim (1-\epsilon) \, {\cal N}\left(\bs 0_p, \begin{pmatrix}  \bs W_1 & \bs 0\\  \bs 0 &  \bs 0 \end{pmatrix}\right) +  \epsilon \, {\cal N}\left(\bs 0_p, \begin{pmatrix}  \bs W_1 & \bs 0\\ \bs 0 & \bs W_2\end{pmatrix}\right),
 \label{model}
\end{equation*}
with $\epsilon <1/2$, $\bs W_1 \in \mathcal{P}_{r_1}$ and $\bs W_2 \in \mathcal{SP}_{p-r_1}$ with $\rank(\bo W_2) = r_2 \leq p-r_1$.\\
Such a distribution illustrates a model containing two clusters: the majority of the data and a group that can be identified as outlying observations. The first cluster follows a Gaussian distribution such that the majority of the data is contained in a $r_1$-dimensional subspace spanned by the range of $\bs W_1$. The observations from the second cluster behave the same as previously on the $r_1$-dimensional subspace but they are also present in $r_2$ directions not spanned by the majority of the data. The goal of the ICS method is to find this $r_2$-dimensional subspace where the observations of the second cluster are outlying. Here this subspace is spanned by the range of $\bo W_2$ which is the orthogonal complement of the range of $\bo W_1$, i.e. the null space of $\bo W_1$. This specific example is a special case of a more general scenario on which we observe independent random variables
$x_i = y_i + z_i$, with independent components $y_i$ and $z_i$ such that $y_i \in  \mathcal{Y}$  and $z_i \in  \mathcal{Z}$ with linear subspaces  $\mathcal{Y}, \mathcal{Z}$ of  $\mathcal{R}^p$ such that  $\mathcal{Y} \cap  \mathcal{Z} = \{0\}$, and $z_i = 0$ with probability $1- \epsilon$, $\epsilon \in (0, 1/2)$.

Let us try to recover this subspace using the ICS method with a theoretical ``perfectly robust'' scatter functional  $\V_1 = \diag(\bs W_1, \bs 0)$ and a theoretical ``non-robust'' scatter functional, the covariance of $\X$, $\V_2= \diag(\bs W_1, \epsilon \bs W_2)$. We have $\V_1 \in  \mathcal{SP}_p$ and $\V_2 \in  \mathcal{SP}_p$, with $\rank(\V_1)=r_{1}<\rank(\V_2)\leq p$. In addition, $\range(\V_1) = \range( \bo W_1)$ and $\range(\V_2) = \range(\bo W_1) \oplus \range( \bo W_2$).

Several of the aforementioned cases arise on this example. First, the intersection of the spaces spanned by the two scatter functionals $\V_1$ and $\V_2$ corresponds to the $r_1$-dimensional subspace spanned by $\bo W_1$, so $r_1$ nonzero eigenvalues should be found. Then, since $\Null(\V_1) -   \Null(\V_2) \neq \{0\}$,  a new direction associated to an $\infty$ eigenvalue should also be analyzed. In fact, this is the one that reveals the outliers. Finally, if $\rank(\V_2) < p$, then the two scatter functionals share a part of their null subspaces. This subspace is not important since it contains no structure. However, we consider this phenomenon in our analysis because it is common in practice that the data is not of full rank. In addition, this feature could also make some algorithms unstable. 

\begin{figure}[h!]
			\includegraphics[keepaspectratio=true,width=7.5cm]{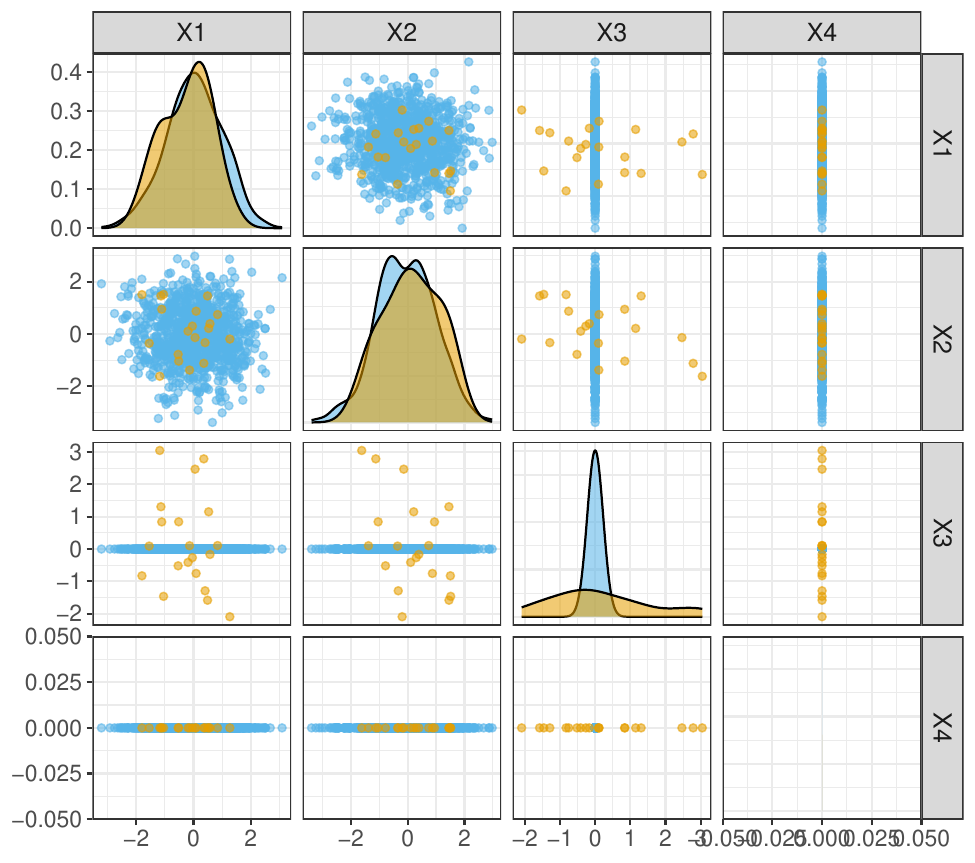} \hfill
			\includegraphics[keepaspectratio=true,width=7.5cm]{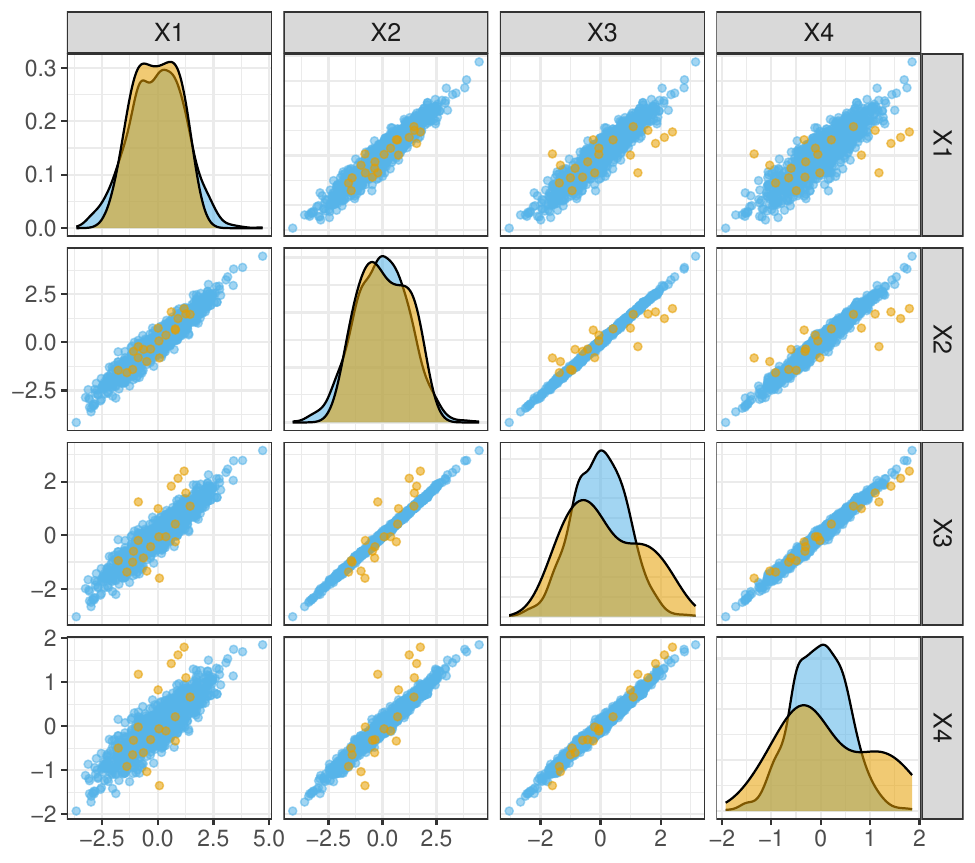}
			\caption{Left: Scatterplot matrix of the simulated observations with 1000 observations, 20 outliers, $\bs W_1=\bs I_2$ and $\bs W_2 = \diag(2,0)$. Right: Scatterplot matrix of the simulated observations transformed by the non-singular matrix $\bo A$.}\label{fig:X_X_G}
	\end{figure}

Practically, to illustrate the model (\ref{model}), we generate 1000 observations with exactly 20 outliers, $\bs W_1=\bs I_2$ and $\bs W_2 = \diag(2,0)$. On the left scatterplot matrix of the Figure~\ref{fig:X_X_G}, we can see that the outliers represented by some blue triangles behave differently than the majority of the data only on the third variable. The subspace spanned by this third variable is the only one of interest to identify these observations as outliers.
This example can be seen as tricky since the outliers are well-identified on the third variable and no observations lie on the fourth one. So, we apply an affine transformation based on a non-singular $p \times p$ particular Toeplitz matrix $\bo A$:
\begin{equation*}\label{eq:toeplitz}
 \bo A = 
\begin{pmatrix}
1 						& \frac{p-1}{p} & \frac{p-2}{p} & \frac{1}{p} \\ 
\frac{p-1}{p} & 1 						& \frac{p-1}{p} & \frac{p-2}{p}\\ 
\frac{p-2}{p} & \frac{p-1}{p} & 1 						& \frac{p-1}{p} \\
\frac{1}{p} 	& \frac{p-2}{p} & \frac{p-1}{p} & 1
\end{pmatrix},
\end{equation*}
 to transform the initial data $\X$ to $\X^* = \bo A \X  $. We can notice on the right scatterplot matrix of Figure~\ref{fig:X_X_G}, that the outliers are no longer as well separated on the third transformed variable as they were initially. In addition, we are no longer able to see that the observations lie in a three-dimensional subspace. However, the structure of outlierness of the data is still contained in one dimension only. The challenge is to be able to recover the direction spanned by the outliers with ICS.

\section{Implementation of ICS for semi-definite positive scatter matrices\label{sec:implementation}}
Several methods exist to solve a Generalized Eigenvalue Problem (GEP) directly, without transforming it into to a non-symmetric EVP. Among these methods are the well-known QZ-algorithm introduced by \cite{moler_algorithm_1973} and the procedure described by \cite{schott_matrix_2005}. However, in practice, solving a GEP of two scatter matrices with a common null space from a numerical point of view is particularly challenging. The presence of this null space makes procedures like the well-known QZ-algorithm very unstable. So far, the available algorithms are not satisfactory as they might lead to complex and negative eigenvalues. 
Additionally, these algorithms can suffer from numerical instability and inaccurate results when dealing with nearly singular matrices. For this reason, GEP is generally not solved directly, and surrogate approaches are used. One common approach is to shift the problem by adding a small multiple $\eta$ of the identity matrix to the (nearly) singular scatter matrix, converting it into a non-singular one.  In practice, this method works only if the matrix is nearly singular or not highly singular. Moreover, the $\eta$ value needs to be tuned, adding an additional step to the procedure. However, the main limitation lies in the assumption that the scatter matrix can be computed from the initial ill-conditioned data, which is usually not possible for robust scatter matrices. A solution involves analyzing regularized scatter matrices instead of the usual scatter matrices, as mentioned in Subsection~\ref{subsec:scatters}. Nonetheless, computing such matrices can be computationally expensive, especially for large-scale problems where the regularized parameters need to be tuned. In addition, depending on the regularization, they might not comply with the definition of a scatter and are not necessarily affine equivariant estimators anymore. \cite{tebbens_improving_2007} review additional methods to solve a GEP in the LDA context, which could also be adapted to ICS.

In this section, we focus on the issue of collinearity when $n>p$ and we investigate three theoretical and practical implementations of ICS based on  Moore-Penrose pseudo-inverse (GSVD) in Subsection~\ref{subsec:MP}, dimension reduction (DR) in Subsection~\ref{subsec:DR} and 
generalized singular value decomposition (GSVD) in Subsection~\ref{subsec:GSVD}. We also compare their properties in terms of: (i)~the criterion to optimize, (ii)~the affine invariance of the scores, and (iii)~the symmetry of the roles of the two scatter estimates $\V_1$ and $\V_2$.

\subsection{ICS with a Moore-Penrose pseudo-inverse}\label{subsec:MP}
If we assume that $\V_1$ and $\V_2$ can be defined and computed on $\X_n$ but that $\V_1 \in \SP$ with $\rank(\V_1)=r_1<p$, then it is not possible to solve the nonsymmetric EVP for the matrix $\V_1^{-1}\V_2$ since $\V_1$ is not invertible. Instead, we can replace the inverse of $\V_1$ with its Moore-Penrose pseudo-inverse $\V_1^+$ and solve the EVP for $\V_1^+ \V_2$:
\begin{equation}
    \V_1^+ \V_2 \bo b = \rho \bo b,
\end{equation}
The pseudo-inverse of  $\bo V_1 \in  \mathcal{SP}_p$ with $\rank(\bo V_1)=r_1<p$ can be defined based on its spectral decomposition: $\bo V_1 = \bo P \bo \Lambda \bo P'$, where $\bo \Lambda$ is a diagonal matrix with decreasing eigenvalues of $\bo V_1$,  $\lambda_1 \geq \dots \lambda_{r_1} > \lambda_{r_1+1} = \dots =\lambda_p=0$ and $\bo P$ is an orthogonal matrix containing the  eigenvectors of $\bo V_1$. $\bo P$ can be partitioned as  $\bo P = [\bo P_1~ \bo P_2]$, with $\bo P_1$, the $p\times r_1$ matrix containing the first $r_1$ eigenvectors associated to the  $r_1$ nonzero eigenvalues of $\bo V_1$, which is an orthonormal basis for the range space of $\bo V_1$. Similarly, the $p\times p-r_1$ matrix $\bo P_2$ spans the null space of $\bo V_1$. $\bo P_1$ and $\bo P_2$  are semi-orthogonal matrices such that:  $\bo P_{r_1}'\bo P_{r_1} = \bo I_{r_1}$ and $\bo P_{p-r_1}'\bo P_{p-r_1} =  \bo I_{p-r_1}$. $\bo V_1$ can be rewritten as $\bo V_1= \bo P_{r_1} \bo \Lambda_{r_1} \bo P_{r_1}'$ and so its pseudo-inverse is:
\begin{equation*}
\bo V_1^+ = \bo P \bo \Lambda^+ \bo P' =\bo P_{r_1} \bo \Lambda_{r_1}^{-1} \bo P_{r_1}',
\end{equation*}
with  $\bo \Lambda_{r_1}^{-1}$ containing only the inverse of the  $r_1$ nonzero eigenvalues of $\bo V_1$.

\begin{propr} Solving the nonsymmetric EVP for the matrix $ \V_1^+ \V_2$ restricts the direction $\bo  b$ associated to the largest eigenvalue $\rho_1$ to $\bo  b = \V_1 +\V_0$ with $\V_1 \in \range(\V_1)$, $\V_0 \in \Null(\V_1)$ only onto the subspace spanned by $\V_1$ and expressed by $\V_1$:
\begin{equation*}
    \V_1=   \ \underset{ \bo b \in \mathcal{R}^p, \bo b \neq 0}{\argmax}	\frac{\bo b^\top \bo \Proj_{\V_1} \V_2\bo \Proj_{\V_1}\bo b}{\bo b^\top \bo P_1 \bo \Lambda_{r_1} \bo P_1^\top \bo b}, 
\end{equation*}
where $\bo \Proj_{\V_1}= \bo P_1 \bo P_1^\top$. The roles of $\V_1$ and  $\V_2$ are not exchangeable anymore.
\end{propr}

\begin{proof}[\textbf{\upshape Proof:}] 
Instead of solving the non-symmetric EVP for the matrix $\V_1^+ \V_2$, we transform it to the symmetric matrix M=$\V_1^{+1/2} \V_2  \V_1^{+1/2}$:
\begin{equation}\label{def:GinvSym}
     \V_1^+ \V_2 \bo b = \rho \bo b \Leftrightarrow
\V_1^{+1/2} \V_2  \V_1^{+1/2} \bo b^* = \rho \bo b^*
  \Leftrightarrow
(\bo \Lambda_{r_1}^{-1/2} \bo P_1^\top\V_2 \bo P_1 \bo \Lambda_{r_1}^{-1/2}-\rho \bo I_{r_1})\bo b^*=0,
\end{equation}

with $\bo b=\bo P_1 \bo \Lambda_{r_1}^{-1/2} \bo b^*$. By multiplying by $\bo P_1\bo \Lambda_{r_1}^{1/2}$, and because  $\bo P_1$ is only semi-orthogonal, the equation (\ref{def:GinvSym}) can be rewritten as:
\begin{equation*}
( \bo P_1 \bo P_1^\top\V_2 \bo P_1  \bo P_1^\top -\rho \bo P_1 \bo \Lambda_{r_1} \bo P_1^\top)\bo b=0,
\end{equation*}
which leads to the following modified ICS criterion for the eigenvector associated with the largest eigenvalue:

\begin{equation}\label{def:ICScriteriumPseudoInv}
 \underset{ \bo b \in \mathcal{R}^p, \bo b \neq 0}{\max} ~~~ \frac{\bo b^\top \bo P_1 \bo P_1^\top\V_2 \bo P_1  \bo P_1^\top  \bo b}{\bo b^\top \bo P_1 \bo \Lambda_{r_1} \bo P_1^\top \bo b}, 
\end{equation}
with $\bo P_1 \bo P_1^\top= \bo \Proj_{\V_1}$, an orthogonal projection matrix onto the $\range(\V_1)$, as $\bo P_1$ is an orthonormal basis for $\range(\V_1)$. So $\range(\bo P_1 \bo P_1^\top\V_2 \bo P_1 \bo P_1^\top) \subseteq \range(\V_1)$ and $\range(\bo P_1 \bo \Lambda_{r_1} \bo P_1^\top) =  \range(\V_1)$. In addition,  $\mathcal{R}^p$ can be decomposed such that: $\mathcal{R}^p = \range(\V_1) \oplus \Null(\V_1)$, thus the solution $\bo b$ of the criterion (\ref{def:ICScriteriumPseudoInv}) can be expressed as:
\begin{equation*}
		\bo  b = \V_1 +\V_0 \text{~~with~~} \V_1 \in \range(\V_1) \text{~~and~~}  \V_0 \in \Null(\V_1), 
\end{equation*}		
\begin{equation*}
		\text{~~with~~}
			\V_1=   \ \underset{ \bo b \in \mathcal{R}^p, \bo b \neq 0}{\argmax}	\frac{\bo b^\top \bo \Proj_{\V_1} \V_2\bo \Proj_{\V_1}\bo b}{\bo b^\top \bo P_1 \bo \Lambda_{r_1} \bo P_1^\top \bo b}. 
\end{equation*}
As $\Null(\V_1) \subseteq \Null( \bo \Proj_{\V_1}\V_2  \bo \Proj_{\V_1})$, optimizing the new criterion (\ref{def:ICScriteriumPseudoInv}) restricts the solutions to directions $\bo b$ only onto the subspace spanned by $\V_1$ and expressed by $\V_1$. 
\end{proof}

\begin{remark}
If the structure of the data is only visible onto the subspace spanned by $\V_2$, in the null space of $\V_1$, then it is not possible to highlight it and recover the outlying observations: if $\bo b \in \Null(\V_1) - \Null(\V_2)$, then  $\rho=\infty$.
The roles of  $\V_1$ and  $\V_2$ are not exchangeable anymore. Indeed, the directions found only span the range of the inverted scatter matrix. So, the ranks of the null spaces of $\V_1$ and  $\V_2$ are now important. The results remain if $\V_2$ is singular or not. However, if $\V_1(\X_n)$ and $\V_2(\X_n)$ are well-defined but potentially (nearly) singular, one idea is to eliminate the common null space of $\V_1$ and $\V_2$ by diagonalizing $(\V_1 + \V_2)^+\V_2$.
\end{remark}

\begin{remark}{Equivalence with the classical ICS.}
If $\V_1 \in \PP$  then solving the $\V_1^{-1} \V_2$ eigenvalue problem or using the Moore-Penroe pseudo-inverse of $\V_1$ is equivalent because $\V_1^+=\V_1^{-1}$.
\end{remark}

\begin{remark}If $\V_2(\X_n)$ is not defined, the standard ICS algorithm is not applicable but computing $\V_2$ on the whitened data might be possible: $\V_2(\X_n \V_1^{+1/2})$.
\end{remark}

\begin{example}
    Going back to our artificial example~\ref{subsec:challenges}, using the Moore-Penrose pseudo-inverse of $\V_1$ leads to optimize the following criterion for finding the first eigenvalue:
\begin{equation*}\label{def:ICScriteriumPseudoInvEx}
 \underset{ \bo b \in \mathcal{R}^p, \bo b \neq 0}{\max} ~~~ \frac{\bo b^\top \bo \Proj_{\bo W_1} \V_2 \bo \Proj_{\bo W_1}   \bo b}{\bo b^\top \bo W_1 \bo b} ~~~ = ~~~ \underset{ \bo b \in \mathcal{R}^p, \bo b \neq 0}{\max} ~~~ \frac{\bo b^\top \bo W_1   \bo b}{\bo b^\top \bo W_1 \bo b}=1.
\end{equation*}
Clearly, in this case, any $\bo b \in \mathcal{R} ^p$ is a solution of the maximization which implies that the structure of outlierness contained in $\bo W_2$ cannot be highlighted. We obtain two eigenvalues equal to one since  $\V_1$ is two-dimensional and two others equal to zero. The projection of the data onto the eigenvectors space is illustrated in Figure~\ref{fig:Z_Ginv}. Definitely, the outliers cannot be identified because the eigenspace is restricted to the subspace spanned by $\V_1$ which does not contain the structure of outlierness defined by $\bo W_2$. So, the pseudo-inverse of $\V_1$ does not always give the correct solution to the singularity issue of the scatter matrices.

\begin{figure}[h!]
			\includegraphics[keepaspectratio=true,width=8.5cm]{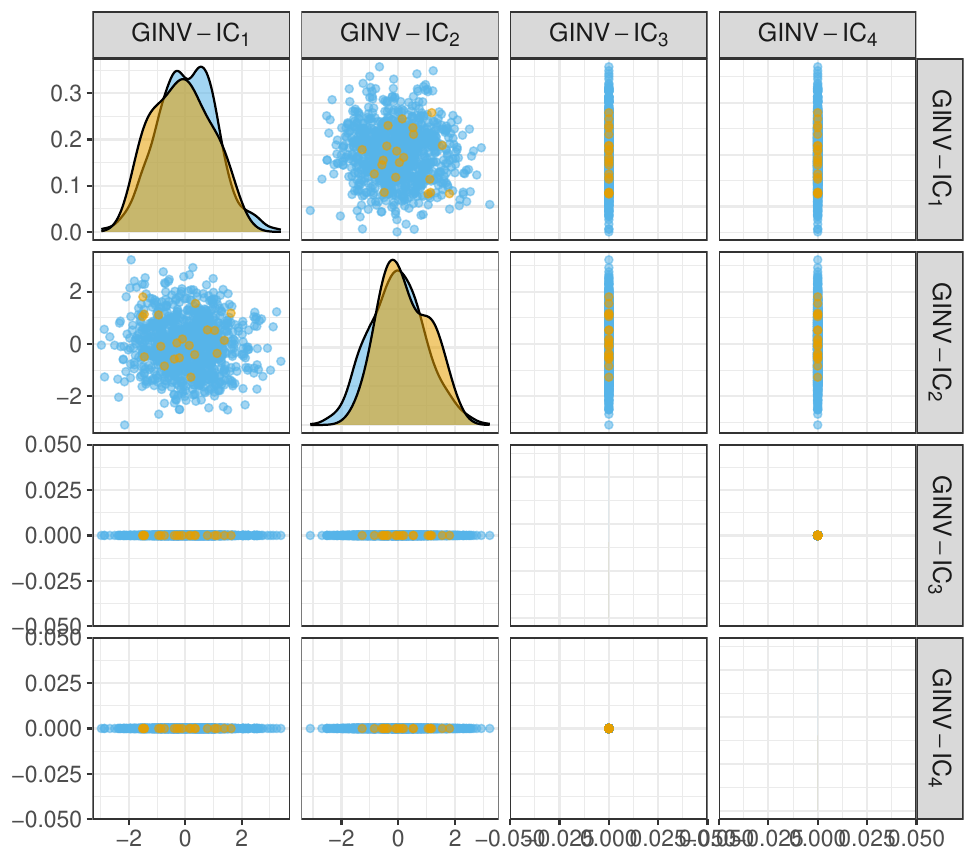} \hfill
			\includegraphics[keepaspectratio=true,width=8.5cm]{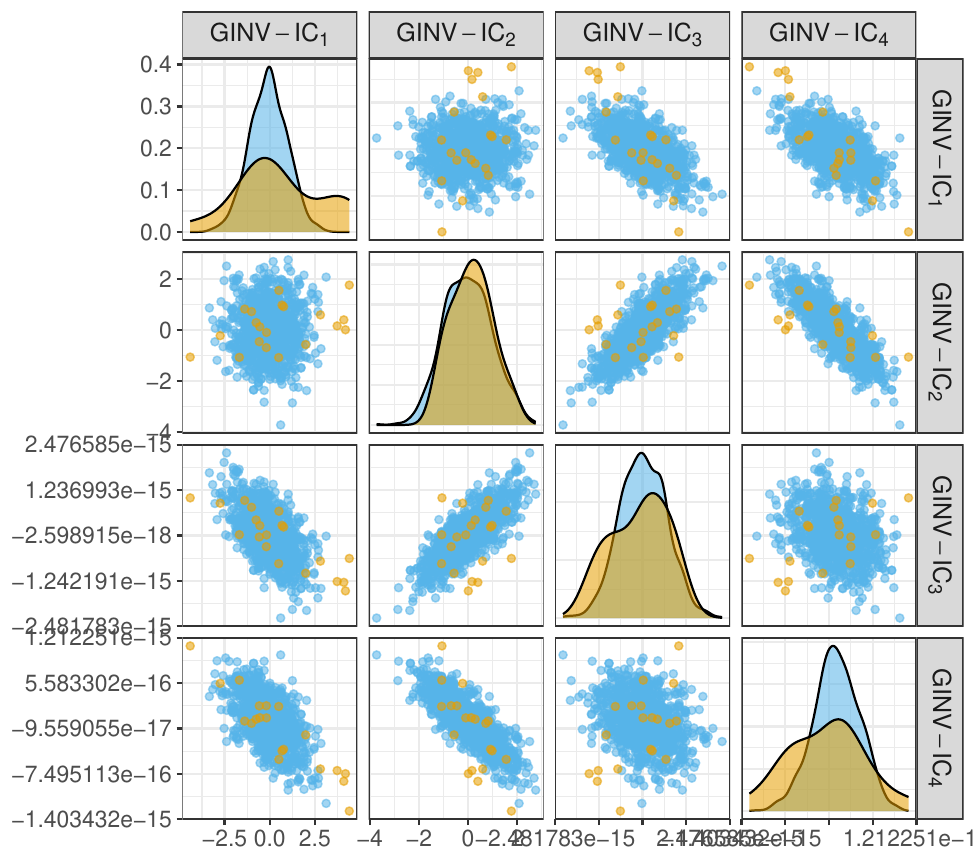}
			\caption{Scatterplot matrix of the IC resulting of ICS using the generalized inverse of $\V_1$ on $\X_n$  (left panel) with $\rho_1=\rho_2=1$ and $\rho_3=\rho_4=0$ and on $\X_n \bo A$  (right panel) with $\rho_1=1.12$, $\rho_2=1$ and $\rho_3=1.4e^{-15}$, $\rho_4=-2.3e^{-16}$.}\label{fig:Z_Ginv}
	\end{figure}

\end{example}

\begin{propr}\label{prop:MP}
If the roots $\rho_1,\dots,\rho_p$ are all distinct, then for the orthogonal transformation $\X_n^* =  \X_n \bo A +\bo 1_n \bo \gamma^\top$, with $\bo A$ being non-singular and $\bs \gamma \in \mathcal{R}^p$, the coordinates
$\Z_n^* =(\X_n^* - \bo 1_n\bs T(\X_n^*)^\top)\bs B({\X_n^*})^\top$ and 
$\Z_n =(\X_n - \bo 1_n\bs T(\X_n)^\top)\bs B({\X_n})^\top$, 
 then
$$ \Z_n^* = \Z_n \bo J,$$
where $\bo J$ is a $p \times p$ diagonal matrix with diagonal elements $\pm 1$, which means the coordinates $\bo Z_n^*$ and $\bo Z_n$  are invariant up to their signs through an orthogonal transformation and not an affine transformation.

\end{propr}
\begin{proof}[\textbf{\upshape Proof:}] 
Let $\X_n^* =  \X_n \bo A + \bo 1_n \bo \gamma^\top$, with $\bo A$ being non-singular and $\bs \gamma \in \mathcal{R}^p$ and $\V_1(\X_n) \in \SP$ with $\rank(\V_1)<p$. By definition of a scatter matrix: $\V_1(\X_n^*) = \bo A^\top \V_1(\X_n) \bo A$ and if  $\bo A$ is an orthogonal matrix, then: 
$\V_1^+(\X_n^*) = \bo A^{-1} \V_1^+(X_n) (\bo A^\top)^{-1} =  \bo A^\top \V_1^+(X_n) \bo A$. 
Following the computations detailed in the proof of Property~\ref{prop:MP}:   $$\V_1^+(\X_n^*) \V_2(\X_n^*) \tilde{ \bo b} = \rho  \tilde{ \bo b}  \Leftrightarrow ( \bo A \bo P_1 \bo P_1^\top \V_2(\X_n) \bo P_1  \bo P_1^\top \bo A^\top -\rho \bo A \bo P_1 \bo \Lambda_{r_1} \bo P_1^\top  \bo A^\top ) \tilde{ \bo b}  =0, $$
with $\bo b=\bo A \bo P_1 \bo \Lambda_{r_1}^{-1/2} \tilde{ \bo b}$ and by multiplying by $\bo A \bo P_1\bo \Lambda_{r_1}^{1/2}$.
This leads to the following modified ICS criterion:
\begin{equation}\label{def:ICScriteriumPseudoInvA}
 \underset{ \bo b \in \mathcal{R}^p, \bo b \neq 0}{\max} ~~~ \frac{\bo b^\top \bo A  \bo \Proj_{\V_1(\X_n)}\V_2(\X_n)  \bo \Proj_{\V_1(\X_n)}   \bo A^\top\bo b}{\bo b^\top  \bo A \bo P_1 \bo \Lambda_{r_1} \bo P_1^\top \bo A^\top \bo b}. 
\end{equation}
Compared to the criterion~(\ref{def:ICScriteriumPseudoInv}), the eigenvectors are rotated by $\bo A^\top$ and so projecting the transformed data $\X_n^*$ onto $\bo B\bo A^\top$ or projecting $\X_n$ onto $\bo B$ leads to the same coordinates $\Z_n^*$ and $\Z_n$ up to their signs. 
\end{proof}

\begin{remark}
If $\V_1 \in \mathcal{SP}_p$ then the ICS coordinates are not necessarily invariant by an affine transformation since the assumption of orthogonality is required in the proof (see the next counter-example~\ref{ctex:Ginv}).
\end{remark}

\begin{example}\label{ctex:Ginv}
We consider the simulated data transformed by the non-singular matrix $\bo A$ (\ref{eq:toeplitz}). In this case, the structure of outlierness of the data is still contained only in one dimension. So, if the two scatter matrices $\V_1$ and $\V_2$ are of full ranks then, doing ICS on the initial data $\X_n$ or on the transformed $\X_n^*=\X_n \bo A$ should lead to the same coordinates. However, if $\rank(\V_1)<p$ and if we use the pseudo-inverse $\V_1^+$ then we lose this affine invariance property of the Invariant Components (IC). Indeed, in the simulated example, we obtain two different eigenvalues, $\rho_1=1.12$ and $\rho_2=1$ instead of the two equal to one and $\rho_3$ and $\rho_4$ are not exactly zero, highligthing some inaccuracies in the computations, as illustrated on the right panel on Figure~\ref{fig:Z_Ginv}. Obviously,  projecting the data onto the eigenvectors' space leads to new scores, and the affine invariance of the coordinates is lost. 

\end{example}

To conclude, using the generalized inverse shows some differences. First, the initial ICS criterion (\ref{def:ICSratio}) may be modified to the criterion (\ref{def:ICScriteriumPseudoInv}), which leads to finding directions only on the subspace spanned by $\V_1$ and so the structure contained in the space spanned by $\Null(\V_1) -\Null(\V_2)$ cannot be highlighted. Second, if we use a generalized inverse, the coordinates are invariant up to an orthogonal transformation as for PCA but no longer to an affine transformation. This is unfortunate since an additional choice is required: standardize the data or not.
Finally, the two scatter functionals $\V_1$ and  $\V_2$ are not exchangeable anymore. Indeed, the directions found only span the range of the inverted scatter matrix. The results remain if $\V_2$ is singular or not.

\subsection{ICS with a dimension reduction as pre-processing}\label{subsec:DR}
Another well-known approach consists of getting rid of the singularity issues by doing a reduction of dimension (DR) first, hoping that no information about the data structure will be lost.  The idea is to perform a Singular Value Decomposition (SVD) of the initial data and to project it onto the right-singular vectors associated with the non-zero singular values. Among others, \cite{hubert_robpca_2005} or \cite{filzmoser_outlier_2008} use this pre-processing step before applying their outlier detection algorithms based on PCA or  Mahalanobis distances. This rank reduction is also used in projection pursuit techniques as mentioned by \cite{hui2010projection} or for the LDA method in the HDLSS context, as explained by \cite{howland_generalizing_2004}. However, the performance of the preprocessing for the LDA method relies on the rank of the covariance matrix which has to fall into a specific range to ensure that the new within-covariance becomes non-singular. Another pitfall is noted by \cite{she_robust_2016} who advised against using an SVD before a robust PCA in the presence of OC outliers.

~

Considering the reduced data  $\X_n^* =\X_n\bo P_1$ of rank $r_{\X_n}$, we have to solve the GEP of $\V_1(\X_n^*)$ and $\V_2(\X_n^*)$:
\begin{equation}\label{GEPReducDim}
\V_2(\X_n^*) \bo b = \rho \V_1(\X_n^*) \bo b,
\end{equation}
with the singular value decomposition of $ \X_n$ such that $ \X_n = \bo U \bo D \bo P^\top$ with $\bo U$ and $\bo P$ two $n ~\times~ n$ and $p ~\times~ p$ orthogonal matrices, $\bo D = \diag(\bo \Delta^{1/2}_{r_{\X_n}\times r_{\X_n}}, \bo 0_{(n-r_{\X_n})\times (p-r_{\X_n})})$0. The elements of the diagonal matrix $ \bo \Delta^{1/2}$ are the square roots of the positive eigenvalues of $\X_n \X_n^\top$ and $\X_n^\top\X_n$. The columns of $\bo P$ are also the eigenvectors of $\X_n^\top\X_n$ and the columns of $\bo U$ are the eigenvectors of $\X_n \X_n^\top$. $\bo U$ and  $\bo P$ can be partitioned as $\bo U=[ \bo U_1 ~\bo U_2]$ where  $\bo U_1$ is $n ~\times~ r_{\X_n}$, $\bo U_2$ is $n ~\times~ (n-r_{\X_n})$ and $\bo P=[ \bo P_1~ \bo P_2]$ where $\bo P_1$ is $p~\times~r_{\X_n}$ and  $\bo P_2$ is $p ~\times~ (p-r_{\X_n})$.  $\bo U _1$ and $\bo P_1$ are both semi-orthogonal matrices and $\bo U_1$ (resp. $\bo P_1$) are orthonormal basis for the column space (resp. the row space) of $\X_n$. 

\begin{remark}If $\X_n$ is of full rank.\\
If $\X_n$ is of full rank then performing an SVD as a preprocessing step before  ICS leads to the same components as if we directly compute the invariant coordinates (IC) from the initial data $\X_n$. Indeed, in this case,  $\rank(\X_n)=r= p$, the data is transformed by a $\bo P$ non-singular orthogonal $p \times p$ matrix and it is known that the invariant coordinates are invariant by an orthogonal transformation. However, from a computational point of view, some numerical discrepancies can arise due to the additional step.
\end{remark}

\begin{remark}If $\range(\V_1(\X_n^*))= \range(\X_n^*)$ but  $\Null(\V_2(\X_n^*)) -   \Null(\X_n^*) \neq \{0\}$.\\
If $\range(\V_1(\X_n^*))= \range(\X_n)$, then  the GEP (\ref{GEPReducDim}) of $\V_1(\X_n^*)$ and $\V_2(\X_n^*)$ can be simplified to the classical EVP: $\V_1(\X_n^*)^{-1} \V_2(\X_n^*) \bo b = \rho  \bo b$.
However,  doing the first step of dimension reduction and transforming the data onto $\X_n^*$ does not ensure the non-singularity of $\V_2(\X_n^*)$. So, it is possible to choose a $\V_2(\X_n^*)$ of lower rank than $\X_n$. In this case, the solution can lead to some directions associated to a zero eigenvalue of multiplicity potentially greater than one.
\end{remark}

\begin{remark}If $\Null(\V_1(\X_n^*)) -   \Null(\X_n^*) \neq \{0\}$.\\
If $\Null(\V_1(\X_n^*)) -   \Null(\X_n^*) \neq \{0\}$, it means that $\V_1(\X_n^*)$ is still singular and doing the first step of dimension reduction does not solve the problem.
\end{remark}

\begin{propr}\label{chap4:prop2SVD}~\\
If $\V_1 =  \cov$ and  $\V_2$ is any scatter matrix as defined in~(\ref{eqt:scatter}), then performing ICS with the Moore-Penrose pseudo-inverse of $\V_1$ or running ICS on the reduced data leads to the same coordinates up to their signs.
\end{propr}

\begin{proof}[\textbf{\upshape Proof:}] 
Let us start with the initial EVP in~\ref{def:ICSvect}:  $\V_2(\X_n) \bo b = \rho   \V_1(\X_n) \bo b$. By multiplying by $\bo P^\top$ and with $\bo b = \bo P \tilde{\bo b}$ the equation can be rewritten as:
\begin{equation*}
 \bo P^\top \V_2(\X_n) \bo P \tilde{\bo b} = \rho \bo P^\top \V_1(\X_n) \bo P \tilde{\bo b} 
 \Leftrightarrow
  \V_2(\X_n \bo P)\tilde{\bo b} = \rho \V_1(\X_n\bo P)  \tilde{\bo b}, 
\end{equation*}

with 
$\V(\X_n \bo P ) = \diag( \V^{(r_{\X_n})}(\X_n \bo P_1), \bo 0_{(p-r_{\X_n}) \times (p-r_{\X_n})})$ and $ \V^{(r_{\X_n})}(\X_n \bo P_1) = \V(\X_n^*)$. Using the notations introduced in  Subsection~\ref{subsec:MP} for the Moore-Penrose pseudo-inverse: $\V_1(\X_n^*)=\bo \Lambda_{r_1}$. This is because the right eigenvectors from the singular value decomposition of $\X_n$ are the ones of $\X_n^\top \X_n=\V_1(\X_n) = \bo P_1 \bo \Lambda_{r_1} \bo P_1^\top$ and $\V_1(\X_n^*)=\bo P_1^\top \V_1(\X_n) \bo P_1$ with $\bo P_1$ being a $p \times r_{\X_n} $ semi-orthognal matrix and so $r_{\X_n}ik<p$. 

Finally, we solve:
\begin{equation*}
\begin{pmatrix}
 \V_2(\X_n^*) & \bo 0_{r_{\X_n}  \times (p-r_{\X_n} )} \\
\bo 0_{(p-r) \times r_{\X_n} } & \bo 0_{(p-r_{\X_n} ) \times (p-r_{\X_n} )}
\end{pmatrix} \tilde{\bo b} =  \rho \begin{pmatrix}
\bo \Lambda_{r_{\X_n} } & \bo 0_{r \times (p-r_{\X_n} )} \\
\bo 0_{(p-r) \times r_{\X_n} } & \bo 0_{(p-r_{\X_n} ) \times (p-r_{\X_n} )}
\end{pmatrix} \tilde{\bo b}, 
\end{equation*}
with $\tilde{\bo b} = \V_1 +\V_0 \text{~~with~~} \V_1 \in \range(\X_n) \text{~~and~~}  \V_0 \in \Null(X_n)$,
and so if we restrict the solutions only onto the subspace spanned by $\range(\X_n^*)$:
\begin{equation*}
\V_2(\X_n^*) \V_1  = \rho \bo \V_1(\X_n^*) \V_1 
\Leftrightarrow
\V_2(\X_n^*) \V_1  = \rho \bo \Lambda_{r_{\X_n} } \V_1 
\Leftrightarrow
       \ \underset{ \V_1 \in \mathcal{R}^p, \V_1 \neq 0}{\argmax}	\frac{\V_1^\top\V_2(\X_n^*)\V_1}{\V_1^\top  \bo \Lambda_{r_1}  \V_1}.
\end{equation*}
If we compare to the modified criterion~\ref{def:ICScriteriumPseudoInv} obtained using the generalized inverse:

\begin{equation*}
 \underset{ \bo a \in \mathcal{R}^p, \bo a \neq 0}{\max} ~~~ \frac{\bo a^\top \bo P_1 \bo P_1^\top\V_2(\X_n) \bo P_1  \bo P_1^\top  \bo a}{\bo a^\top \bo P_1 \bo \Lambda_{r_1} \bo P_1^\top \bo a} 
 \Leftrightarrow
  \underset{ \bo a \in \mathcal{R}^p, \bo a \neq 0}{\max} ~~~ \frac{\bo a^\top \bo P_1 \V_2(\X_n^*) \bo P_1^\top  \bo a}{\bo a^\top \bo P_1 \V_1(\X_n^*) \bo P_1^\top \bo a}. 
\end{equation*}
with $\bo a= \bo a_1 +\bo a_0$ with $\bo a_1 \in \range(\V_1(\X_n)) =  \range(\X_n)$ and $\bo a_0 \in \Null(\V_1(\X_n))$. So, after projecting, the new coordinates are the same up to their signs. 
\end{proof}

\begin{remark}
So in this case, doing the pre-processing leads to an additional step to the method which is not needed and which implies the same drawbacks as doing ICS with a generalized inverse.
\end{remark}

\begin{remark}
 From a practical point of view, estimating the rank of $\X_n$ might be very challenging as illustrated in Subsection~\ref{subsec:estim_rank} and lead to a loss of information regarding the structure of the data.
\end{remark}

To conclude, the preprocessing step of dimension reduction does not fulfill all its promises. First, it cannot guarantee that it solves the singularity issues of the scatter matrices. Then, even if it does, if we choose $\V_1$ as the variance-covariance matrix, we recover exactly the same modified criterion to solve as when we use the generalized inverse and so the same drawbacks. Finally, if we choose $\V_2$ as the variance-covariance matrix, we might be unable to recover the structure of the data if it is only contained on the subspace spanned by $\V_1$.

\subsection{ICS with a generalized singular value decomposition}\label{subsec:GSVD}
In this section, we focus on an implementation based on a Generalized Singular Value Decomposition (GSVD) as proposed by 
\cite{howland_structure_2003, howland_solving_2006, howland_generalizing_2004} and \cite{kim_dimension_2005} in the LDA context. More specifically,  they use a GSVD for computing eigenvectors to define the Fisher's discriminant subspace, when the between $\bs \Sigma_B$  and the within-group $\bs \Sigma_W$ covariance matrices, are susceptible to be singular. The only requirement with this method is to express $\bs \Sigma_B$ and $\bs \Sigma_W$ as cross-product matrices, which is easily obtained by their definition. 
This procedure, which uses a GSVD to solve a GEP, can be applied to other scatter matrices which can be expressed as crossproducts.
However, defining a stable algorithm for the GSVD is very challenging and a lot of research was done regarding this topic, such as \cite{paige_towards_1981, paige_computing_1986, bai_csd_1992, bai_computing_1993, bai_new_1993,  golub_matrix_1996} among others. In this section, we present the GSVD procedure as it is given in \cite[Section 75-11]{hogben_handbook_2006}, restricted to the case of real matrices. We retain this definition since it is already implemented into LAPACK and can be used directly in \proglang{R} through the \pkg{geigen} \citep{hasselman_geigen_2019} package.

Let us define $\X_{\V_1} \in \mathcal{R}^{n \times p}$ s.t. $\X_{\V_1}^\top \X_{\V_1} =\V_1=\V_1(\bo X_n)$ and $\bo X_{\bo {V_2}} \in \mathcal{R}^{n \times p}$ s.t. $\bo X_{\bo {V_2}}^\top \bo X_{\bo {V_2}} =\bo {V_2}=\V_2(\bo X_n)$. $\V_1, \V_2   \in \mathcal{SP}_p$  with $\rank(\X_{\V_1}) = \rank(\V_1) = r_1 \leq p$ and $\rank(\X_{\V_2}) = \rank(\V_2) = r_2 \leq p$. Contrary to the two previous methods which solve the GEP (\ref{GEP}): $\V_2 \bo b_i = \rho_i \V_1 \bo b_i$, for $i=1,\dots,p$, we propose to solve the slightly modified following problem for $i=1,\dots,p$:
\begin{equation}\label{def:GEPGSVD}
\beta_i^2  \V_2 \bo b_i = \alpha_i^2 \V_1 \bo b_i \\\Leftrightarrow 
\beta_i^2  \X_{\V_2}^\top \X_{\V_2}  \bo b_i = \alpha_i^2 \X_{\V_1}^\top \X_{\V_1}  \bo b_i 
\end{equation}
where $ \rho_i  =\alpha_i^2/\beta_i^2$ is real, nonnegative and possibly infinite.
The rows of $\bo B$ are the eigenvectors of $\X_{\V_2}^\top\X_{\V_2}-\rho \X_{\V_1}^\top\X_{\V_1}$ or equivalently of $\V_2 -\rho \V_1$, and the ``nontrivial'' eigenvalues are the squares of the generalized singular values: $\rho_i = \alpha_i^2/\beta_i^2$, for $i=p-r+1,\dots,p$, with $r= \rank([\X_{\V_1}^\top, \X_{\V_2}^\top]^\top)$. The ``trivial'' eigenvalues are those corresponding to the leading $p -r$ rows of $\bo B$, which span the common null space of $\X_{\V_1}^\top\X_{\V_1}$ and $\X_{\V_2}^\top\X_{\V_2}$. These eigenvalues are not well defined and are not of interest. All the cases of interest are summarized in Table~\ref{tab:GSVD}. 

\begin{table}[ht]
\begin{center}
		\begin{tabular}{l|lll|c}
\multicolumn{1}{c}{~} & \multicolumn{1}{c}{~} & \multicolumn{1}{c}{~} & \multicolumn{1}{c}{~} & \multicolumn{1}{c}{Optimization of}\\
\multicolumn{1}{c}{Direction} & \multicolumn{1}{c}{$\alpha_i$} & \multicolumn{1}{c}{$\beta_i$} & \multicolumn{1}{c}{$\rho$} & \multicolumn{1}{c}{the ICS ratio (\ref{def:ICSratio})}\\
			\hline
			$\bo b_i \in \range(\V_1) \cap  \range(\V_2)$ & $ \alpha_i \neq 0$ & $\beta_i \neq 0$ & $\rho_i \in \mathcal{R}^{+*}$ & Minimize or maximize\\
			$\bo b_i \in \Null(\V_2) -   \Null(\V_1)$ &$ \alpha_i = 0$& $\beta_i \in \mathcal{R}^{+*}$& $\rho_i=0$& Minimize\\
			$\bo b_i \in \Null(\V_1) -   \Null(\V_2)$ &$ \alpha_i \in \mathcal{R}^{+*}$& $\beta_i =0$ & $\rho_i=\infty$ & Maximize\\
			\hline
		\end{tabular}
	\caption{Summary of the different possible directions $\bo b_i$ depending on the values of $\alpha_i$ and $\beta_i$.}
	\label{tab:GSVD}
 \end{center}
\end{table}

To find the generalized eigenvalues and eigenvectors of the pencil $\X_{\V_2}^\top\X_{\V_2}-\rho \X_{\V_1}^\top\X_{\V_1}$, we can leverage the GSVD technique. Indeed, the Generalized (or Quotient) Singular Value Decomposition (GSVD or QSVD) of two $n \times p$ matrices $\bo X_{\V_1}$ and $\bo X_{\V_2}$ is given by the pair of factorization as:
\begin{equation*} \label{GSVDdef}
\bo X_{\V_1} = \bo U \bo D_1 [ \bo 0 ~\bo R] \bo Q' \text{~~~ and ~~~ } \bo X_{\V_2} = \V \bo D_2 [ \bo 0 ~\bo R] \bo Q'
\end{equation*}
where $\bo U$ and $\V$ are $n \times n$, $\bo Q$ is $p \times p$,  $\bo U$, $\V$ and $\bo Q$ are orthogonal. $\bo R$ is $r \times r$, upper triangular and nonsingular. $[ \bo 0 ~\bo R]$ is $r \times p$ (in other words, the $\bo 0$ is an $r\times (p - r )$ zero matrix). $\bo D_1$ and $\bo D_2$ are $n \times r$. Both are real, nonnegative, and diagonal, satisfying $\bo D_1^\top\bo D_1 +\bo D_2^\top\bo D_2=\bo I_r$.  So, we can rewrite the pair of factorization as:
\begin{equation*} \label{GSVDdef2}
 \X_{\V_1}^\top\X_{\V_1} = \V_1 = \bo B^{-T}  \begin{pmatrix} \bo 0 & \bo 0\\ \bo 0 & \bo D_1^\top \bo D_1\end{pmatrix} \bo B^{-1}\text{~~~ and ~~~ }  \X_{\V_2}^\top\X_{\V_2} = \V_2 = \bo B^{-T}  \begin{pmatrix} \bo 0 & \bo 0\\ \bo 0 & \bo D_2^\top \bo D_2\end{pmatrix} \bo B^{-1}
\end{equation*}
with $\bo B^\top = \bo Q \diag(\bo I_{n-r}, \bo R^{-1})$., or equivalentely:
\begin{equation*}
\bo B \X_{\V_1}^\top\X_{\V_1} \bo B^\top = \bo B \V_1(\X_n) \bo B^\top = \begin{pmatrix} \bo 0 & \bo 0\\ \bo 0 & \bo D_1^\top \bo D_1\end{pmatrix}
~~~\text{and}~~~ 
\bo B \X_{\V_2}^\top\X_{\V_2} \bo B^\top = \bo B \V_2(\X_n) \bo B^\top = \begin{pmatrix} \bo 0 & \bo 0\\ \bo 0 & \bo D_2^\top \bo D_2\end{pmatrix},
		\end{equation*}
Write $\bo D_1^\top\bo D_1 = \diag(\alpha_1^2,\dots, \alpha_r^2) $ and $\bo D_2^\top\bo D_2 = \diag(\beta_1^2,\dots, \beta_r^2)$, the ratios $\alpha_j/\beta_j$ for $j = 1, \dots, r$ are called the generalized singular values. Finally, the generalized eigenvalues of the pencil $\X_{\V_2}^\top\X_{\V_2}-\rho \X_{\V_1}^\top\X_{\V_1}$ are  $ \rho_i  =\alpha_i^2/\beta_i^2$ which are real, nonnegative and possibly infinite and the corresponding eigenvectors are the rows of $\bo B$. Note that the normalization is not the same as the one presented in the standard definition~(\ref{def:ICSvect}) but it can easily be adapted.

Re-writting the GEP (\ref{GEP}) as in~(\ref{def:GEPGSVD})  presents some advantages compared to the other two methods. First, it allows to find all the directions which can reveal some structure of the data in the general case where  $\V_1 \in  \mathcal{SP}_p$ and  $\V_2 \in  \mathcal{SP}_p$, as summarized in the Table~\ref{tab:GSVD}. Second, it is clear that $\V_1$ and $\V_2$ play a symmetric role. This is important since the other methods can miss the structure of the data if it is contained into the subspace spanned by $\V_2$ and in the null space of
$\V_1$ in particular. Third, this formulation is still equivalent to the classical EVP~(\ref{def:ICSvect}) if the two scatter matrices are of full ranks with $\rho_i=\alpha_i^2/\beta_i^2$. In addition to these nice characteristics,  the invariant coordinates remain invariant by affine transformation.

\begin{propr}\label{chap4:propGSVD}\textbf{Affine invariance property.}\\
For two affine equivariant scatter matrices $\V_1$ and $\V_2$, and using the eigenvectors defined by~(\ref{def:GEPGSVD}), the invariant coordinates are invariant by an affine transformation.
\end{propr}

\begin{proof}[\textbf{\upshape Proof:}] 
Adaptation of the proofs from \cite{tyler_invariant_2009}, appendix A.1, for distinct and multiple roots detailed in Appendix~\ref{append:GSVD_aff}.
\end{proof}

\begin{example} 
Let us take the same example as previously from Subsection~\ref{subsec:challenges}. Using the GSVD of $\X_{\V_1}$ and $\X_{\V_2}$ to solve the GEP (\ref{def:GEPGSVD}) leads to investigate four different cases for the direction $\bo b$ as illustrated in the left panel of Figure~\ref{fig:Z_GSVD}:
\begin{itemize}[noitemsep,topsep=0pt,parsep=0pt,partopsep=0pt]
	\item if $\bo b \in \range(\V_1) \cap  \range(\V_2) = \range(\bo W_1)$, then the direction $\bo b$ is restricted to the subspace spans by $\bo W_1$ as when we use  the Moore-Penrose pseudo-inverse and we obtain two eigenvalues equal to one,
	\item if $\bo b \in \Null(\V_2) -   \Null(\V_1) = \{0\}$, then no direction $\bo b$ exists,
	\item if $\bo b \in \Null(\V_1) -   \Null(\V_2) = \range(\bo W_2)$, then  $\rho=\infty$ because $\beta^2=0$ and so the direction $\bo b$ can highlight the structure of outlierness contained into the $\range(\bo W_2)$, which is not the case when we use the  Moore-Penrose pseudo-inverse,
	\item	if $\bo b \in \Null(\V_1) \cap  \Null(\V_2) = \Null(\V_2)$, then $\rho$ is a ``trivial'' eigenvalue and any direction $\bo b \in \mathcal{R}^p$ is a solution.
\end{itemize}
However, only the ``non-trivial'' eigenvalues, corresponding to the first three cases, are interesting for highlighting the structure of the data. More precisely, in this example, only the eigenvector $\bo b \in \Null(\V_1) -   \Null(\V_2) = \range(\bo W_2)$ associated with the infinite eigenvalue, contains the structure of outlierness of the data. This is clearly visible in Figure~\ref{fig:Z_GSVD} which illustrates the projection of our simulated data onto the eigenvectors space. So, using the GSVD outperforms the use of a Moore-Penrose pseudo-inverse because it recovers the structure of outlierness of the data. In addition, in the right panel of Figure~\ref{fig:Z_GSVD}, we can see we obtain the same coordinates and eigenvalues if we transform the initial data by the non-singular matrix $\bo A$.

\begin{figure}[h!]
			\includegraphics[keepaspectratio=true,width=8.5cm]{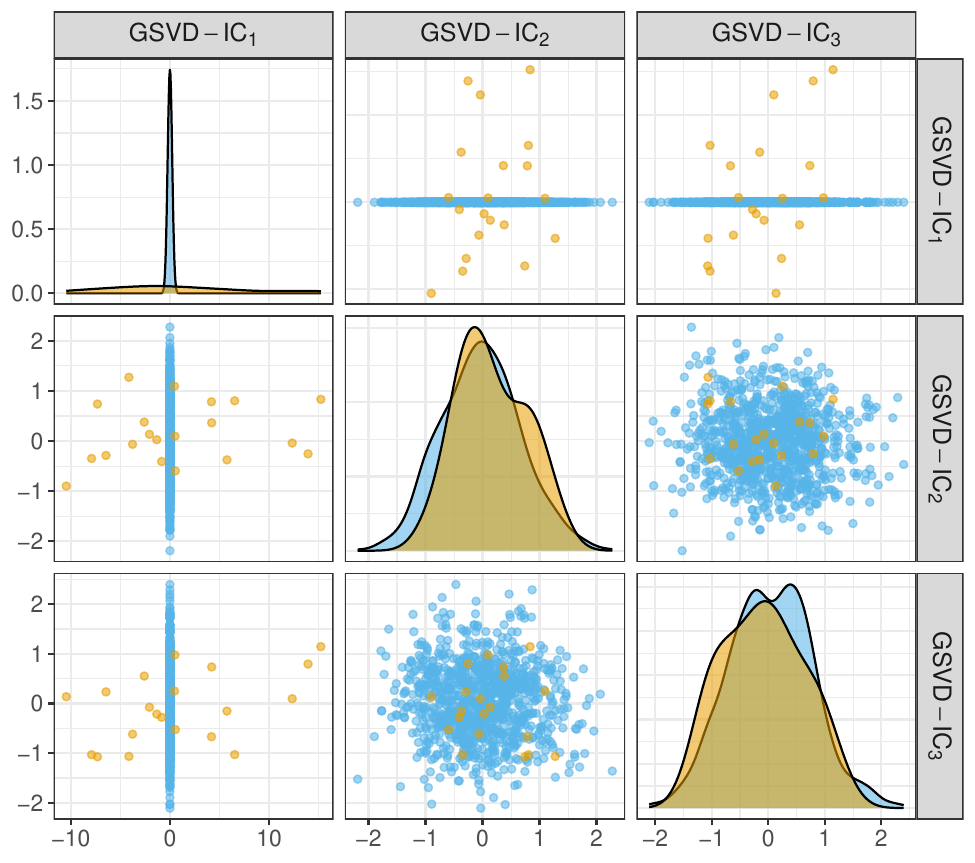}\hfill
			\includegraphics[keepaspectratio=true,width=8.5cm]{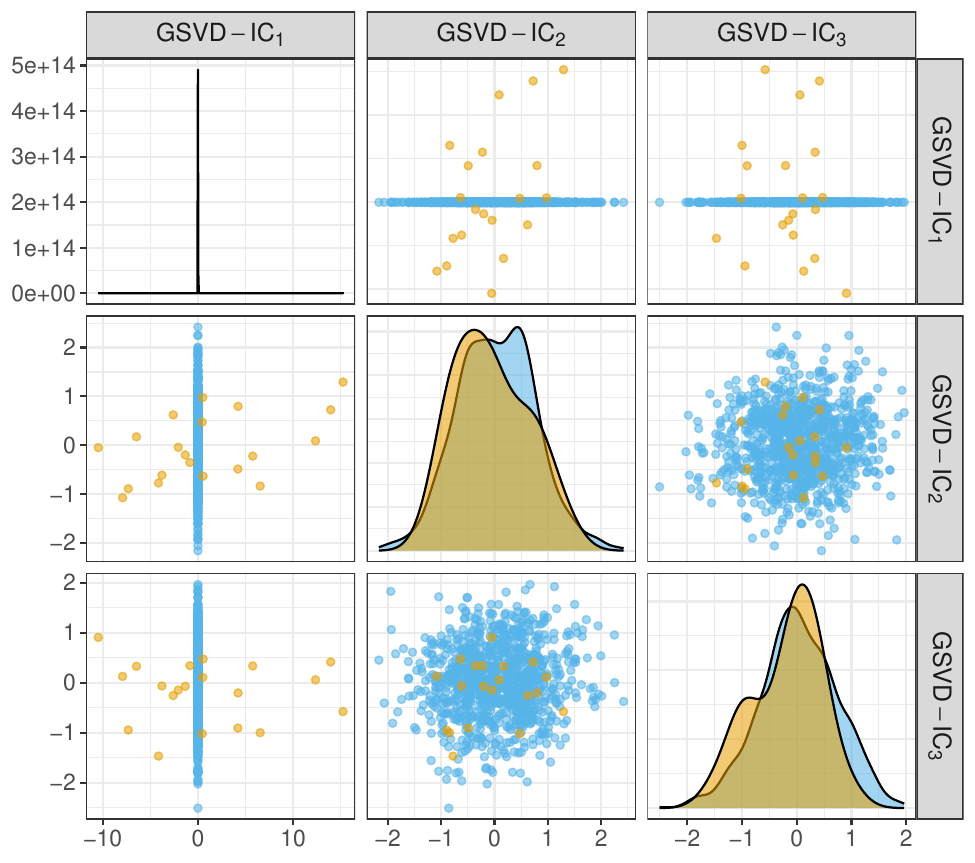}
			\caption{Scatterplot matrix of the IC resulting of ICS using the GSVD of $\X_{\V_1}$ and $\X_{\V_2} $ on $\X_n$ (right panel)  with $\rho_1=\infty$, $\rho_2=\rho_3=1$  and on transformed data $\X_n \bo A$ (right panel)  with $\rho_1=\infty$, $\rho_2=\rho_3=1$.}\label{fig:Z_GSVD}
	\end{figure}	

\end{example}

To conclude, solving the GEP of $\V_1$ and $\V_2$ through the GSVD of $\X_{\V_1}$ and $\X_{\V_2}$ presents three major advantages. First, it solves the possible singularity issues of $\V_1$ and/or $\V_2$ by searching in all directions, and remains equivalent to the EVP of $\V_1^{-1} \V_2$ if the scatter matrices are of full ranks. In addition, it leads that $\V_1$ and $\V_2$ play a symmetric role since the symmetry of the problem is kept. The affine invariance property of the scores continues to be valid in the general case of semi-definite positive scatter matrices. Finally, it is  noteworthy that this GSVD procedure is already implemented in the \pkg{geigen} R package and has proven to be stable. By focusing on the GSVD of $\X_{\V_1}$ and $\X_{\V_2}$, more accurate results can be achieved while reducing the computational burden. This approach eliminates the need for the complete calculation of  $\V_1$ and $\V_2$. However, in practice, it is difficult to define another affine equivariant scatter estimator than the variance-covariance matrix.

Overall, all three approaches present some advantages and limits as summarized in Table~\ref{tab:all}. The next section investigates if in practice those properties are kept.

\begin{table}[h!]
    \centering
    \begin{tabular}{p{1.5cm}p{1.5cm}p{1.5cm}p{3cm}p{2.1cm}p{4.5cm}}
    \hline
        Methods   & Invariance & Symmetry of $\V_1$ and $\V_2$ & Directions found  & Implementation stability& Limits \\
      \hline
      Variable selection   & ? & ?  & $\subseteq \range(\X_n^{selected})$ & Stable: $\sim$  & Challenging to find the subspace of interest of non-collinear variables. \\
      Regularized scatters   & No  & ?  & $\subseteq \range(\tilde{\V_1(\X_n)})$ & Stable: $\sim$  & Might be challenging to compute with additional tuning parameters. Not a scatter strictly census. \\
      Generalized inverse   & Ortho. & No & $\subseteq \range(\V_1(\X_n))$  &  Stable: $\sim$ & Rank estimation is challenging. Do not solve the issue of computing $\V_2(\X_n)$.\\
      Dimension reduction   &  Ortho.  & No & $\subseteq \range(\V^{(r_{\X_n})}(\X_n^*))$ & QR\cite{archimbaud_numerical_2023}: stable & Rank estimation is challenging. QR\cite{archimbaud_numerical_2023} estimation is limited to one-step M-scatter matrices. Do not ensure that $\V_1(\X_n^*)$ is not singular.\\
      GSVD   & Affine & Yes & All & Stable & Only for scatter matrices which can be expressed as crossproducts.\\
      GEP   & Affine &  Yes & All & Unstable & Can lead to negative and complex eigenvalues.\\
      \hline
    \end{tabular}
    \caption{Summary of the main characteristics of the different approaches for solving ICS when at least one scatter is singular. Ortho. stands for orthogonal invariance. $?$  indicates that it depends on the situation and $\sim$ that it depends on the algorithm.}
    \label{tab:all}
\end{table}

\section{Empirical applications\label{sec:applications}}
In this section, we illustrate the characteristics of the different approaches on different empirical applications. We restrict our analysis to ICS with a generalized inverse (GINV), pre-processed by a dimension reduction (DR) through a singular value decomposition and using the generalized value decomposition (GSVD). We exclude the direct GEP approach since it might result in complex and negative eigenvalues. First, Subsection~\ref{subsec:simlpe_col} analyses the consequences of exchanging the role of $\V_1$ and $\V_2$ on a simulated correlated mixture of Gaussian distributions. Subsection~\ref{subsec:estim_rank} focuses on two examples for which estimating the rank of the data is challenging. Subsection~\ref{subsec:HTP2} evaluates the impact of transforming the data through an affine transformation on a collinear industrial data set.  Finally, Subsection~\ref{subsec:HDLSS} investigates if those approaches are applicable also in case of high dimension low sample size (HDLSS) with $n<p$.

\subsection{Comparison of the three approaches: collinear clustering application}\label{subsec:simlpe_col}
Let $\bo X = (\X_1, \dots, \X_d)^\top$ be a $d$-variate real random vector distributed according to a mixture of two Gaussian distributions such that:
\begin{equation}\label{mixt}
{\bo X} \sim \epsilon_1 \, {\cal N}(\bo \mu_1, \bo I_d) +  \epsilon_2 \, {\cal N}(\bo \mu_2, \bo I_d) ,
\end{equation}
with $\epsilon_1 + \epsilon_2 =1$, $\bo \mu_{1} = \bo 0_d$, $\bo \mu_{2} = (\delta, 0, \dots, 0)^\top$  where $\delta=10$. 
We generate $n=1000$ observations on $d=3$ variables for two balanced groups with $\epsilon_1 = \epsilon_2 =0.5$. Two collinear variables are added: $\X_4 = \X_2-3\X_3$ and $\X_5 = \X_3 +5\X_4$.

\subsubsection{$\cov$-$\cov_4$}
\begin{figure}[h!]
			\includegraphics[keepaspectratio=true,width=5cm]{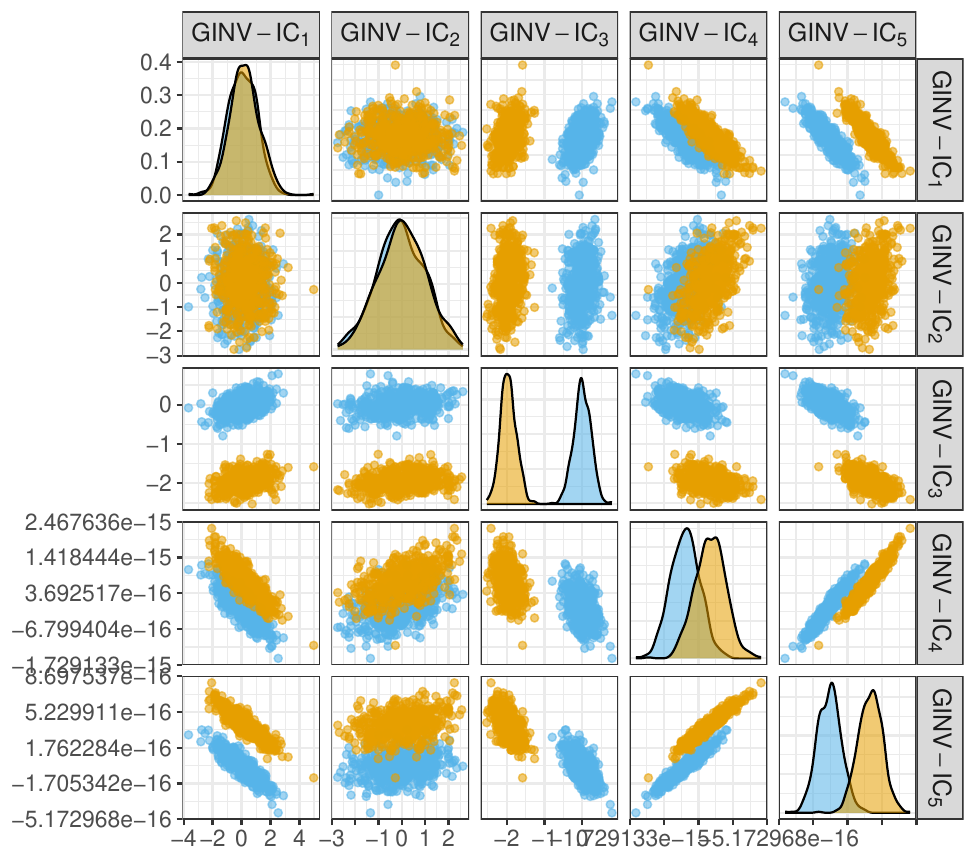} \hfill
			\includegraphics[keepaspectratio=true,width=5cm]{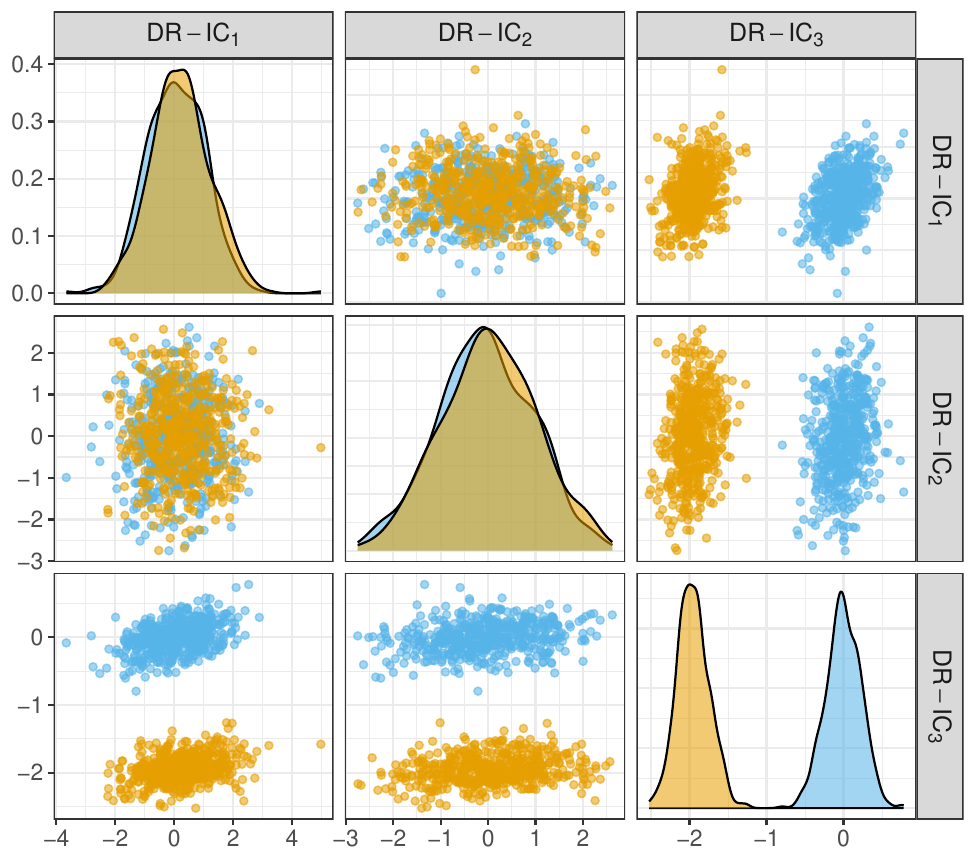}\hfill
			\includegraphics[keepaspectratio=true,width=5cm]{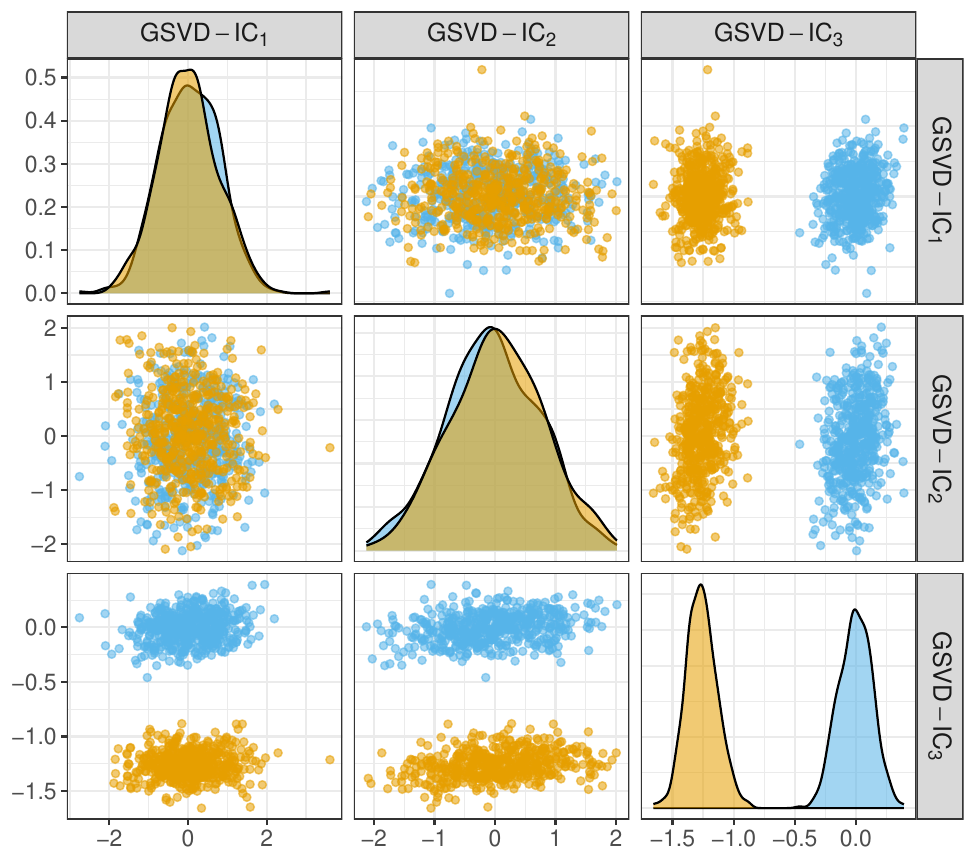}\vfill
   	\includegraphics[keepaspectratio=true,width=5cm]{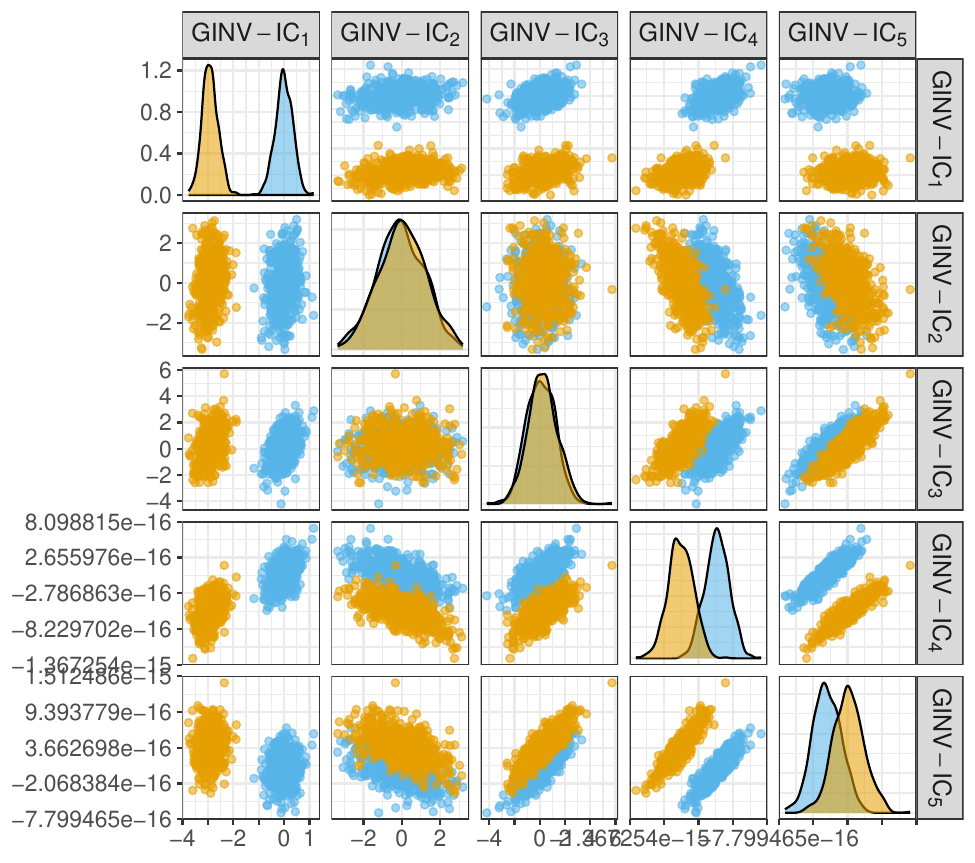} \hfill
			\includegraphics[keepaspectratio=true,width=5cm]{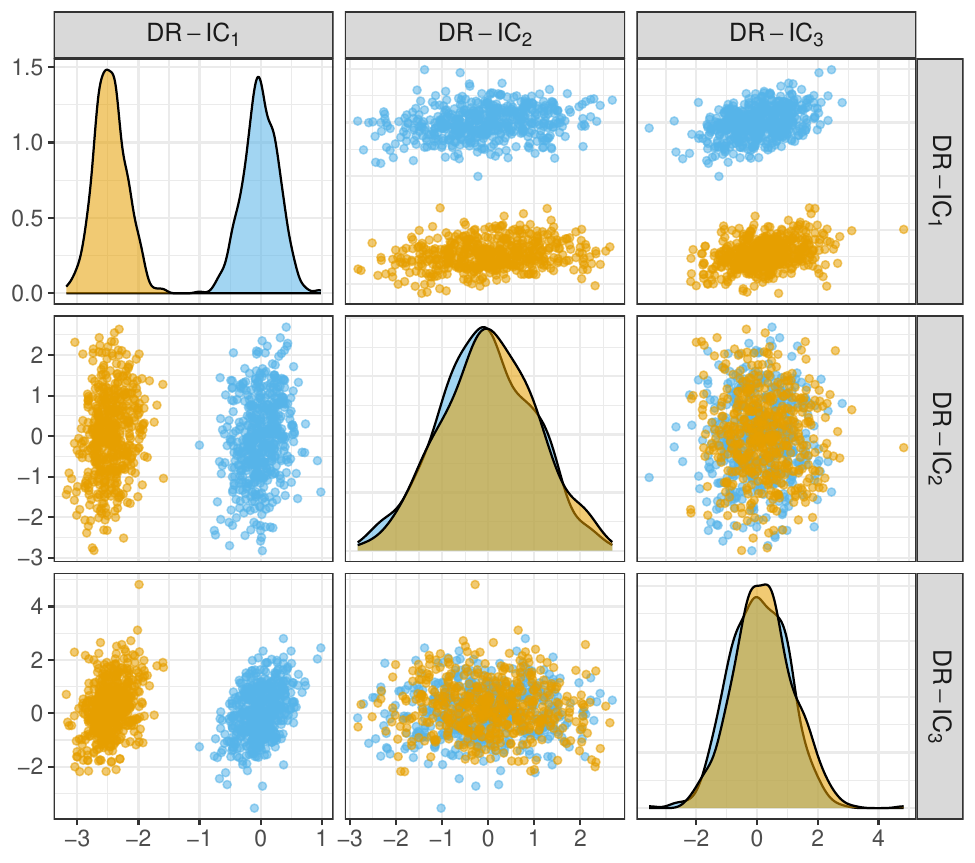}\hfill
			\includegraphics[keepaspectratio=true,width=5cm]{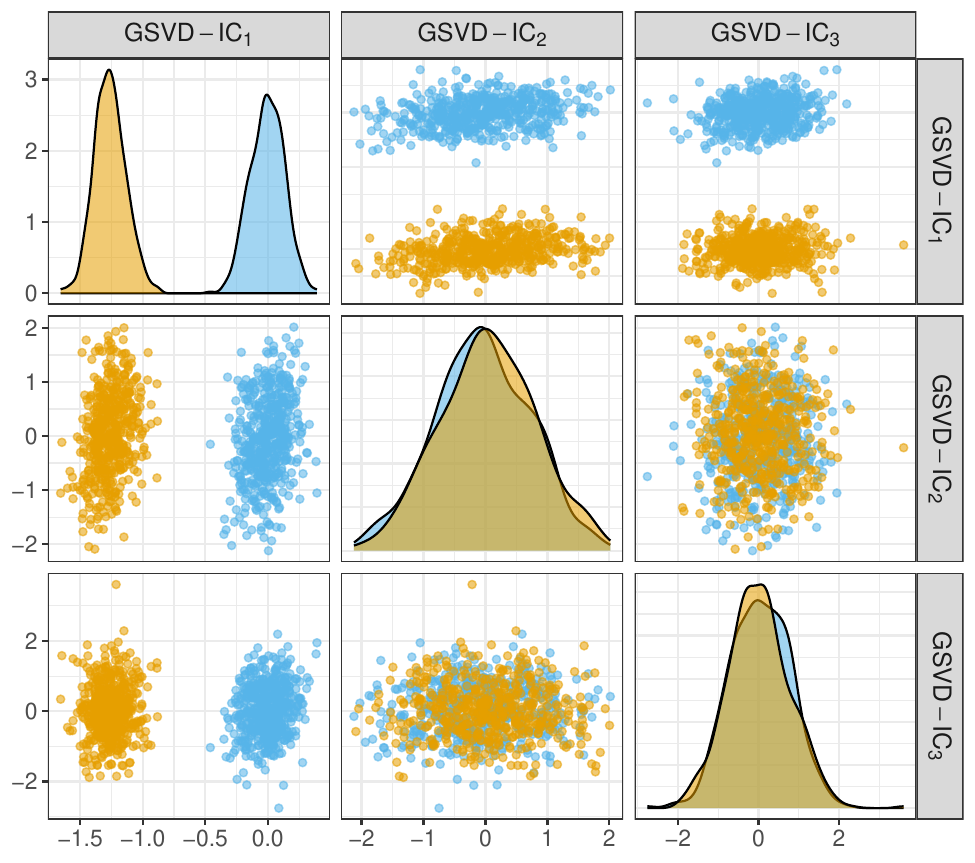}
			\caption{Scatterplots matrix of the IC resulting of ICS of $\V_1$ and $\V_2$ using the generalized inverse of $\cov$ (GINV on 1$^{st}$ column), after a dimension reduction (DR on  2$^{nd}$ column) and with a generalized singular value decomposition (GSVD on 3$^{rd}$ column). The second row illustrates the same results when we exchange $\V_1$ and $\V_2$.}\label{fig:simple_col}
\end{figure}

First of all, we compare the different ICS methods for the scatter pair $\cov-\cov_4$. For GSVD, we consider $\covg_4$ since it is not possible to compute $\cov_4(\X_n)$ as $\X_n$ is not full rank and so $\cov(\X_n)$ is singular and it cannot be inverted. Figure~\ref{fig:simple_col} shows the scatterplots matrix of the IC resulting of ICS with the generalized inverse of $\cov$ (GINV on 1$^{st}$ column), after a dimension reduction (DR on  2$^{nd}$ column) and with a generalized singular value decomposition (GSVD on 3$^{rd}$ column). The second row illustrates the same results when we exchange $\V_1$ and $\V_2$.

It is interesting to note that depending on the method we do not obtain the same number of components: 5 with GINV and 3 for the other two. Indeed, with a DR the rank of the data is estimated to be 3 and so only three dimensions are kept. For GSVD, three non-trivial eigenvalues are also detected. For GINV, 5 components are illustrated but the last two are associated with almost zero eigenvalues: $1.5e^{-15}$ and $-2.5e^{-18}$. Clearly, this indicates numerical issues and the last two components should be disregarded. Now, if we focus on the first three, we can notice that all the methods allow us to easily identify the two clusters on IC$_3$. In addition, the components are the same between GINV and DR as mentioned in Property~\ref{chap4:prop2SVD}. Finally, if we exchange $\V_1$ and $\V_2$ then the clustering structure is shown on the first component for the three methods. So here, on an example of simple collinearity it appears that the three methods lead to similar results. The only point of attention is with GINV, where some trivial eigenvalues are estimated and should be put away.

\subsubsection{$\mcd_{0.5}$-$\cov$}
\begin{figure}[h!]
			\includegraphics[keepaspectratio=true,width=5cm]{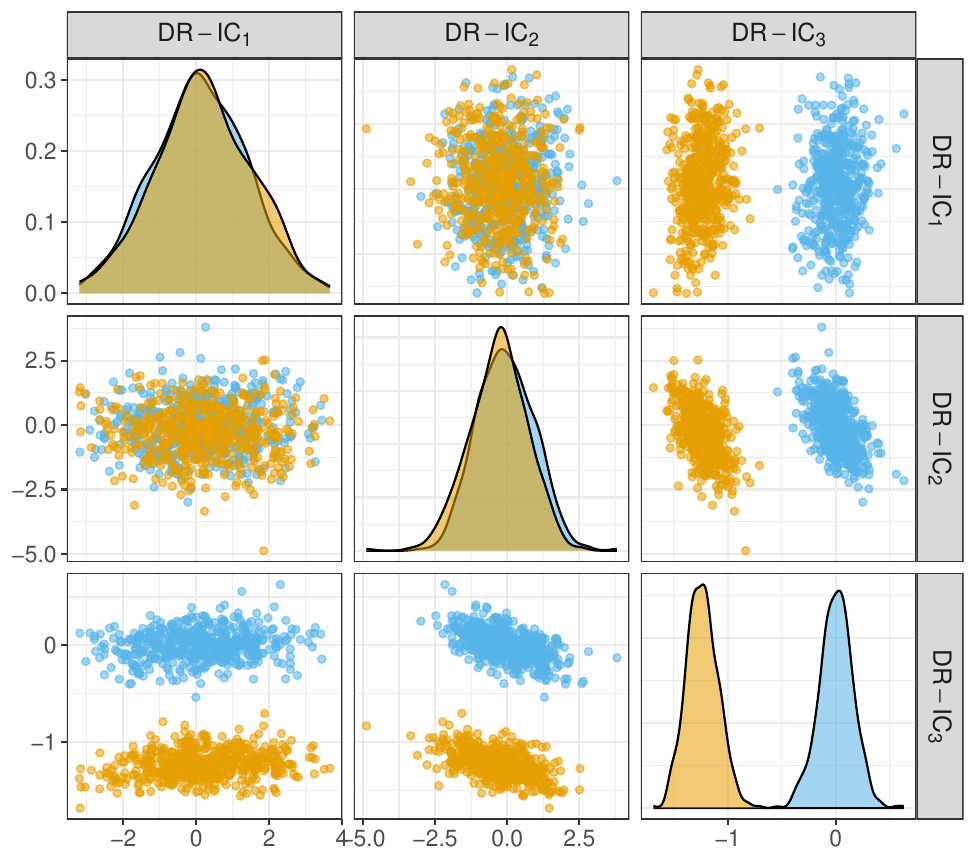} \hfill
			\includegraphics[keepaspectratio=true,width=5cm]{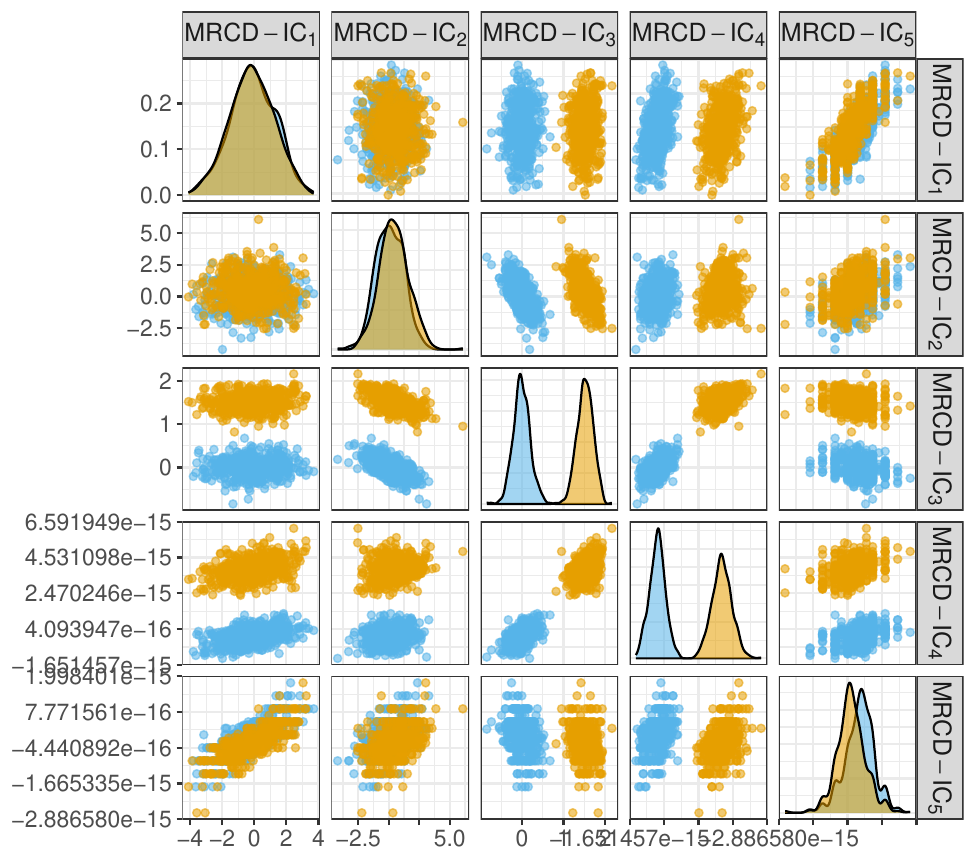}\hfill
			\includegraphics[keepaspectratio=true,width=5cm]{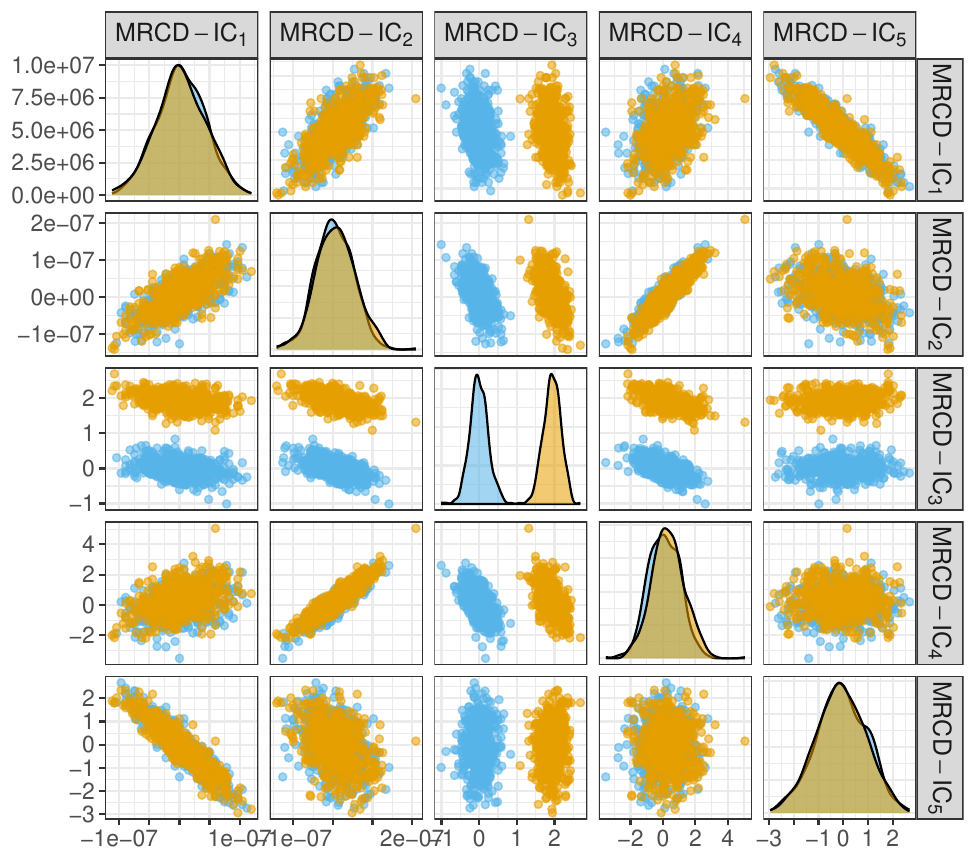}
			\caption{Scatterplots matrix of the IC resulting of ICS of $\mcd_{0.5}-\cov$ after a dimension reduction (DR on  1$^{st}$ column) and of $\mrcd_{0.5}-\cov$ (on 2$^{nd}$ column) and of $\cov-\mrcd_{0.5}$ (on 3$^{rd}$ column).}\label{fig:simple_col_mcd}
\end{figure}

Focusing on a different scatter pair based on a more robust scatter matrix such as the $\mcd_{0.5}$ raises several issues. Indeed, it is not possible to compute the $\mcd_{0.5}$ on our data because ``More than half of the observations lie on a hyperplane'', so GINV and GSVD are not applicable. Instead, we can perform ICS with the $\mcd$ on the reduced data or use its regularized version $\mrcd_{0.5}$ as shown in Figure~\ref{fig:simple_col_mcd}. As previously, with the DR approach, only three dimensions are kept and the clusters are identifiable on the third IC. This is also true with the $\mrcd$ but five eigenvalues are estimated for $\mrcd_{0.5}-\cov$ or $\cov-\mrcd_{0.5}$. In addition, two of those eigenvalues are really small or high: $3.5e^{-15}$, $-5.3e^{-15}$ and $6.7e^{+14}$, $1.9e^{+14}$. So in each case with $\mrcd_{0.5}$ some eigenvalues need to be disregarded and it is an additional step to take into account.  

\subsection{Challenging estimation of the rank}\label{subsec:estim_rank}
This section investigates the difficulty of correctly estimating the rank of two empirical applications: a nearly singular industrial data in Subsection~\ref{subsec:nearly_col} and some simulated data with OC outliers in Subsection~\ref{subsec:OC}.

\subsubsection{HTP3: nearly singular industrial data}\label{subsec:nearly_col}
We consider the HTP3 data set, analyzed by \citep{archimbaud_numerical_2023} and available in the \proglang{R} package \pkg{ICSOutlier}\cite{nordhausen_icsoutlier_2023}. It describes $n=371$ high-tech parts designed for consumer products and characterized by $p=33$ tests. The part 32 showed defects in use and is considered as an outlier. Here, the data set contains tests in different units which lead to nearly singularity and so the classical ICS algorithm returns an error. 

\begin{figure}[h!]
			\includegraphics[keepaspectratio=true,width=5cm]{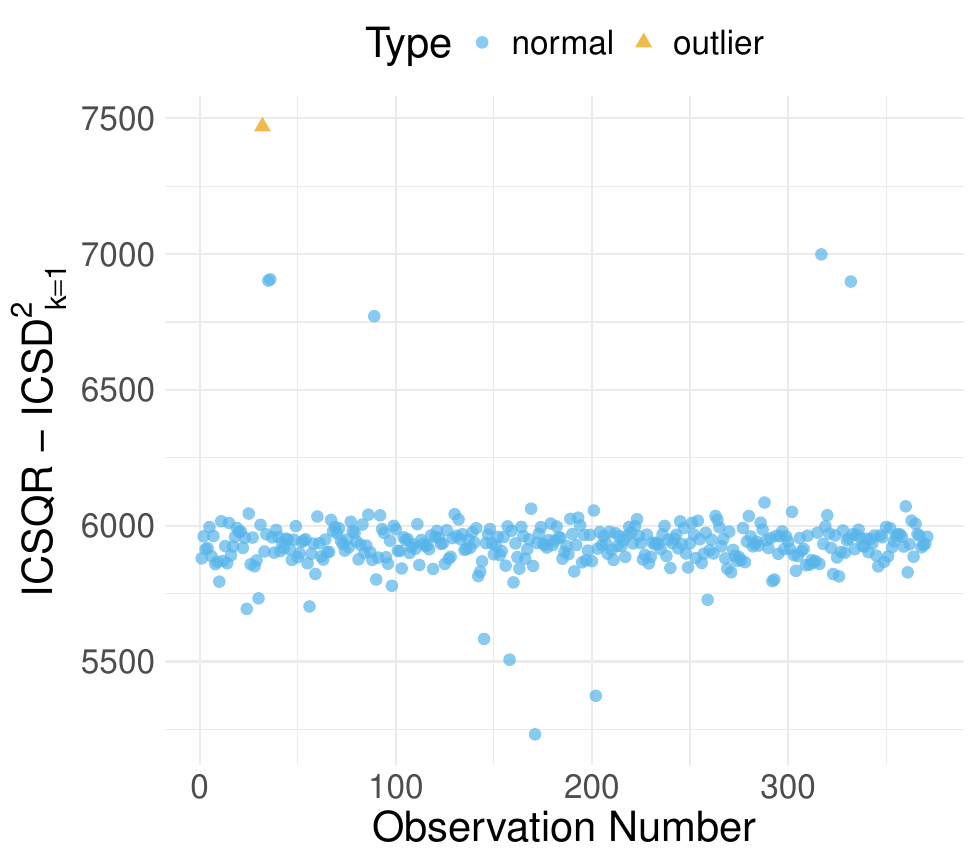} \hfill
			\includegraphics[keepaspectratio=true,width=5cm]{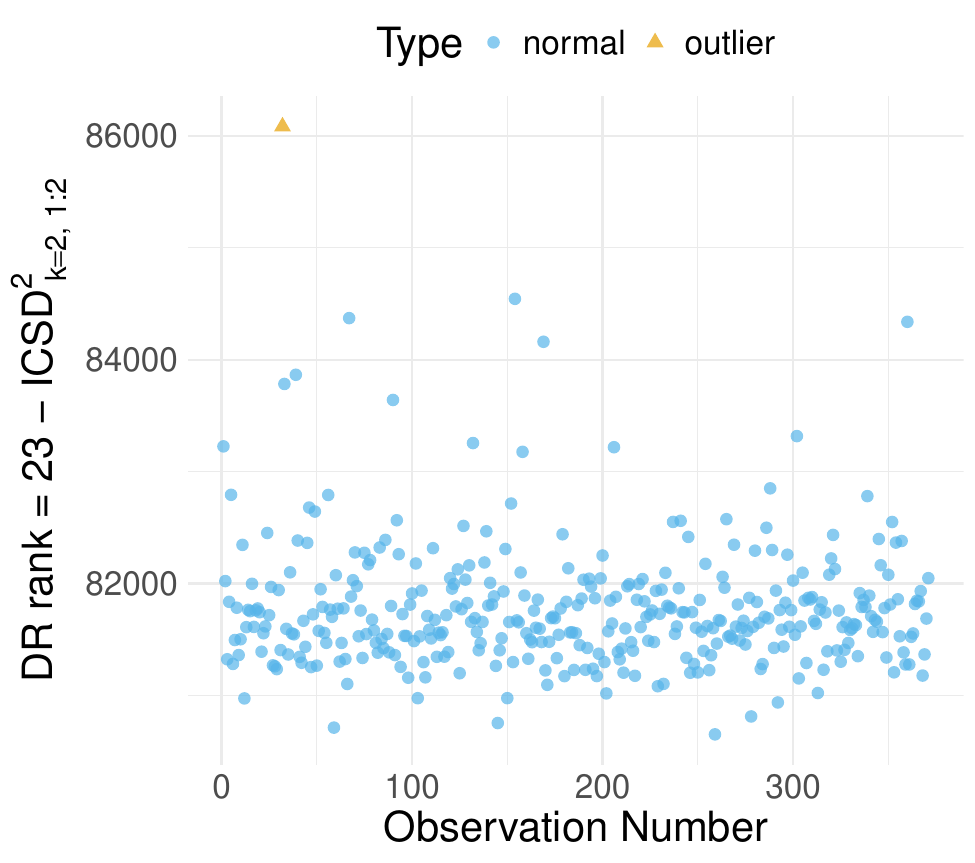}\hfill
			\includegraphics[keepaspectratio=true,width=5cm]{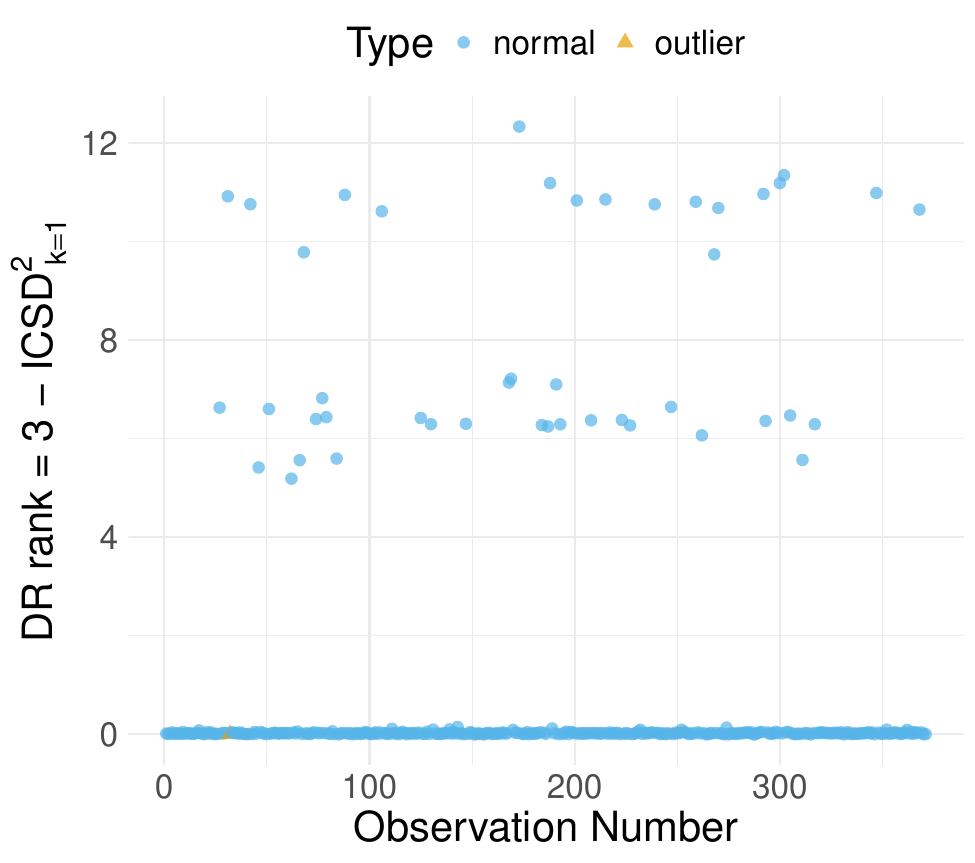}
			\caption{HTP3 data set: ICS distances, ICSD$^2$, computed with $k$ components for ICS with QR algorithm (1$^{st}$ panel), after DR with $\rank(\X_n) = 23$ (2$^{nd}$ panel) and with $\rank(\X_n) = 3$ (3$^{rd}$ panel). The defective part is  in orange.}\label{fig:htp3}
\end{figure}

Using the ICSQR implementation presented in~\cite{archimbaud_numerical_2023} solves the issue for a combination of scatter based on a one-step M-scatter matrix and $\cov$. In Figure~\ref{fig:htp3}, we compute the so-called squared ICS distances~\cite{archimbaud_ics_2018}, denoted ICSD$^2$, of the $k$ selected components for different approaches to perform ICS. On the first plot, we use ICSQR with only the first component and the defective part (in orange) is clearly identified as having a high distance and so being an outlier compared to the other observations. We obtain similar results if we use the GSVD approach with $\cov-\cov_4$ or $\cov_4-\cov$ as illustrated in Appendix~\ref{fig:htp3_append}. However, GINV is not working in this context as we can never compute $\cov_4$. For the DR approach, a new challenge arises regarding the estimation of the rank of the SVD. As mentioned in~\cite{archimbaud_numerical_2023}, it is common practice to use a relative rule to estimate the rank based on the first eigenvalue $\lambda_i$ with $i=1,\dots,p$ such as: (i)  $\lambda_i/\lambda_1 < \sqrt(\nu)$ with the epsilon machine $\nu = 2.2^{-16}$, (ii) $\lambda_i/\lambda_1 < \max(n,p)\nu$ or (iii) $(\sum_{i=1}^l\lambda_i^2)/(\sum_{i=1}^p\lambda_i^2) < 0.99$, to explain at least 99\% of the inertia as with PCA for example. Here, we obtain respectively a rank of 23, 33 or 3 based on the different criteria, meaning that we do not reduce the dimension in the second case. For the others, as we can see in the second column of Figure~\ref{fig:htp3}, the defective part is identified with $\rank=23$ only if we take two components, and it is not detectable in case of $\rank=3$. Considering different scatter pairs as the $\mcd_{0.5}$ is tricky because the scatter matrix cannot be computed on the reduced data. In this case, it is necessary to consider $\cov-\mcd_{0.5}$ and not $\mcd_{0.5}-\cov$, but the results are not improved as visible in the Appendix~\ref{fig:htp3_append}. So on real datasets, the estimation of the rank for the DR approach can be very challenging and not the best approach.

\subsubsection{OC outliers}\label{subsec:OC}
This issue for estimating the rank is even more problematic in case of the presence of OC outliers. We generate $n=100$ observations following the projected mean-shift outlier model presented by \citep{she_robust_2016} without noise: $\X_n = \bo U \bo D \V^\top +(1\bo \mu^\top+S)\V^\top$ with random orthogonal $\bo U$, $\V$, $p=5$, $r=3$, $\bo D = \diag\{1000, 400, 200\}$, $\bo \mu=0$, and the row outlier matrix $\bo S$ has first $O$ rows as $L*[1,\dots,1]$ and $\bo 0$ otherwise, $O=4$ and $L=3.5$.

\begin{figure}[h!]
			\includegraphics[keepaspectratio=true,width=5cm]{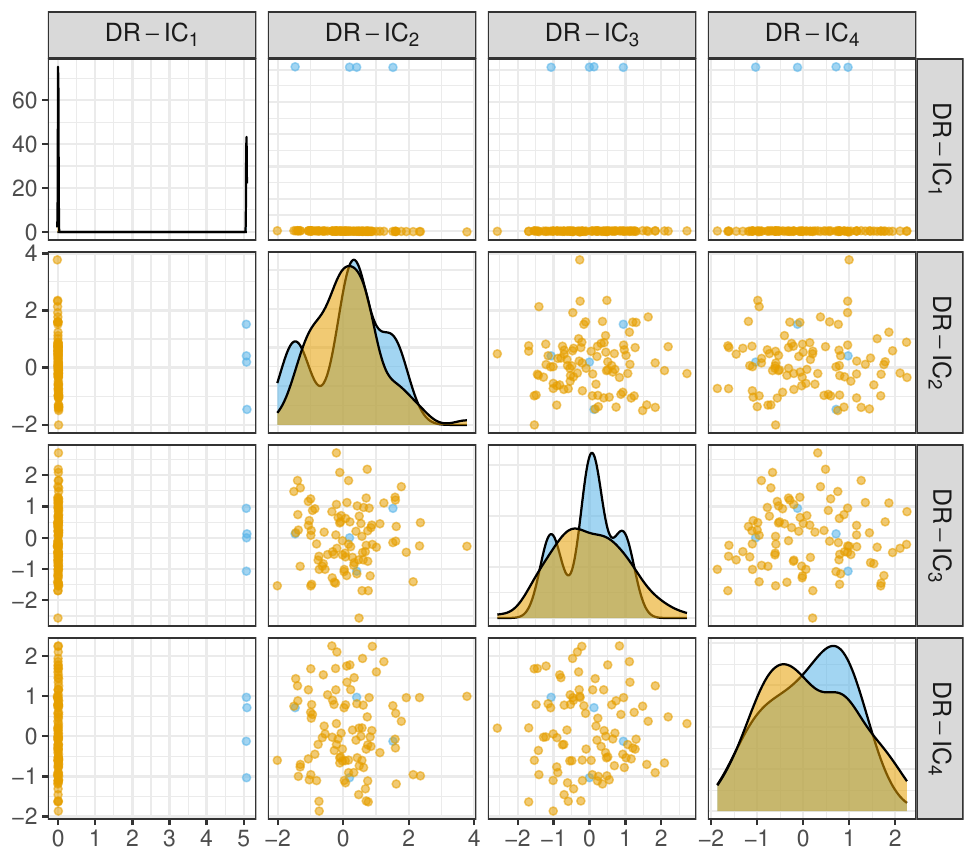} \hfill
			\includegraphics[keepaspectratio=true,width=5cm]{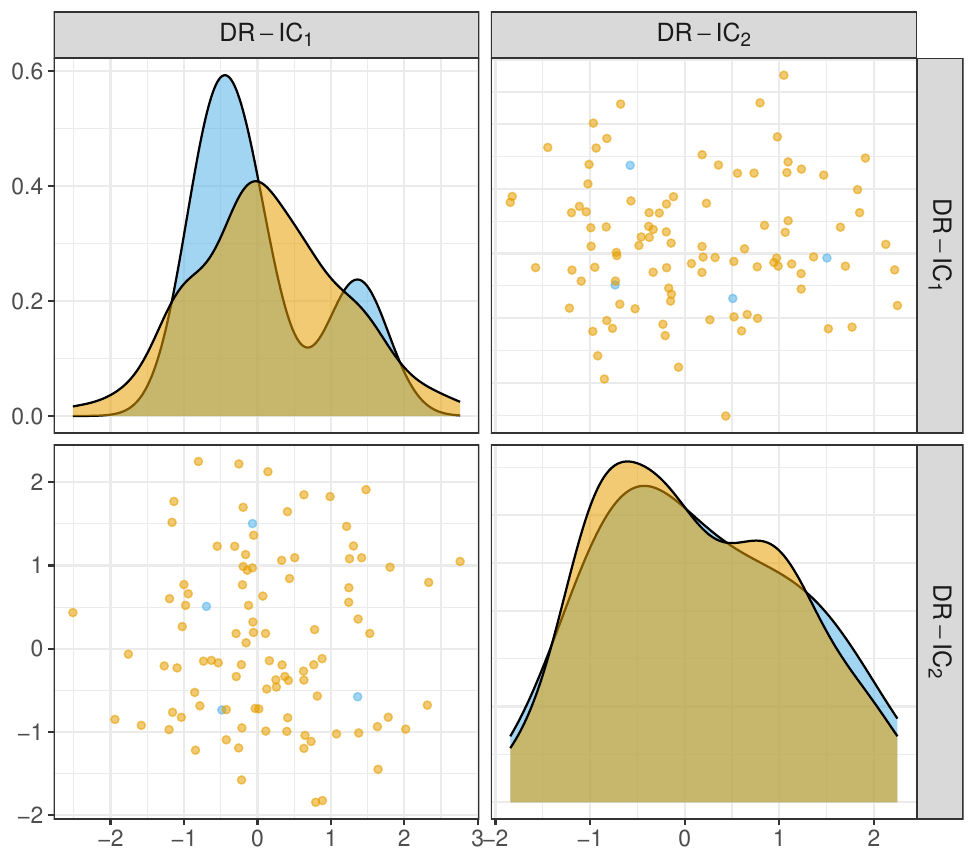}\hfill
			\includegraphics[keepaspectratio=true,width=5cm]{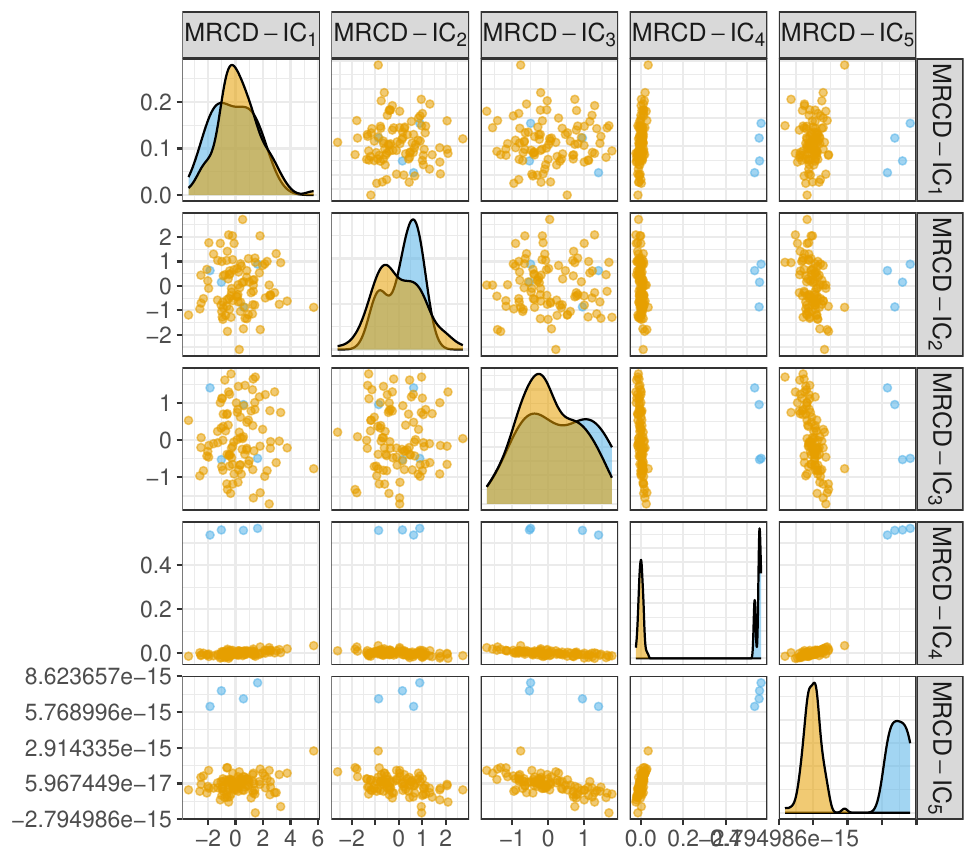}
			\caption{OC outliers: scatterplots matrix of the IC resulting of ICS of $\cov4-\cov$ after a dimension reduction with $\rank=4$ (on  1$^{st}$ column), $\rank=2$ (on 2$^{nd}$ column) and of $\mrcd_{0.5}-\cov$ (on 3$^{rd}$ column).}\label{fig:OC_5}
\end{figure}

In this context, the true rank of the data set is equal to 4 and so the classical ICS returns an error. With a DR step first, if we estimate the rank to 4 then ICS with $\cov-\cov4$ identifies the outliers on IC$_1$ as illustrated in Figure~\ref{fig:OC_5}. However, it is not possible to compute a more robust scatter matrix like the $\mcd_{0.5}$. In this context, we can estimate the rank based on 95\% of explained variance, leading to two dimensions but the information about the OC outliers is lost no matter the scatter pair. However, using a GSVD or GINV approach directly works fine as visible in the second and third plots of Figure~\ref{fig:oc_append} in Appendix~\ref{append:applications}. With $\mrcd-\cov$ the situation is a bit tricky because the outliers are found on IC$_2$ as illustrated in Figure~\ref{fig:OC_5}.

\subsection{Impact of affine transformation: HTP2 - collinear industrial data}\label{subsec:HTP2}
We consider another industrial data set also analyzed by \citep{archimbaud_numerical_2023} and available in the \proglang{R} package \pkg{ICSOutlier} \cite{nordhausen_icsoutlier_2023} called HTP2 to evaluate the impact of an affine transformation on the data such as the classical standardization. It contains $n=149$ tests for $p=457$ high-tech parts with a defective part at number 28. This data set is ill-conditioned and so the classical ICS algorithm returns an error as well as the ICSQR implementation and the GINV approach.

\begin{figure}[h!]
			\includegraphics[keepaspectratio=true,width=5cm]{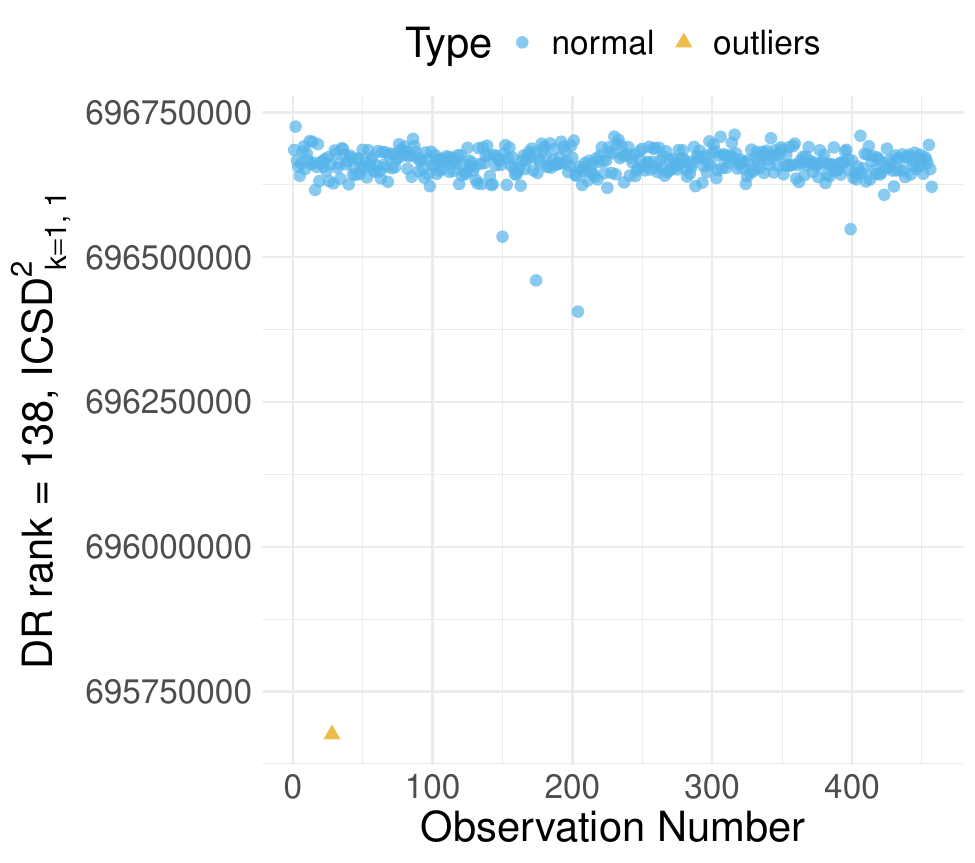} \hfill
			\includegraphics[keepaspectratio=true,width=5cm]{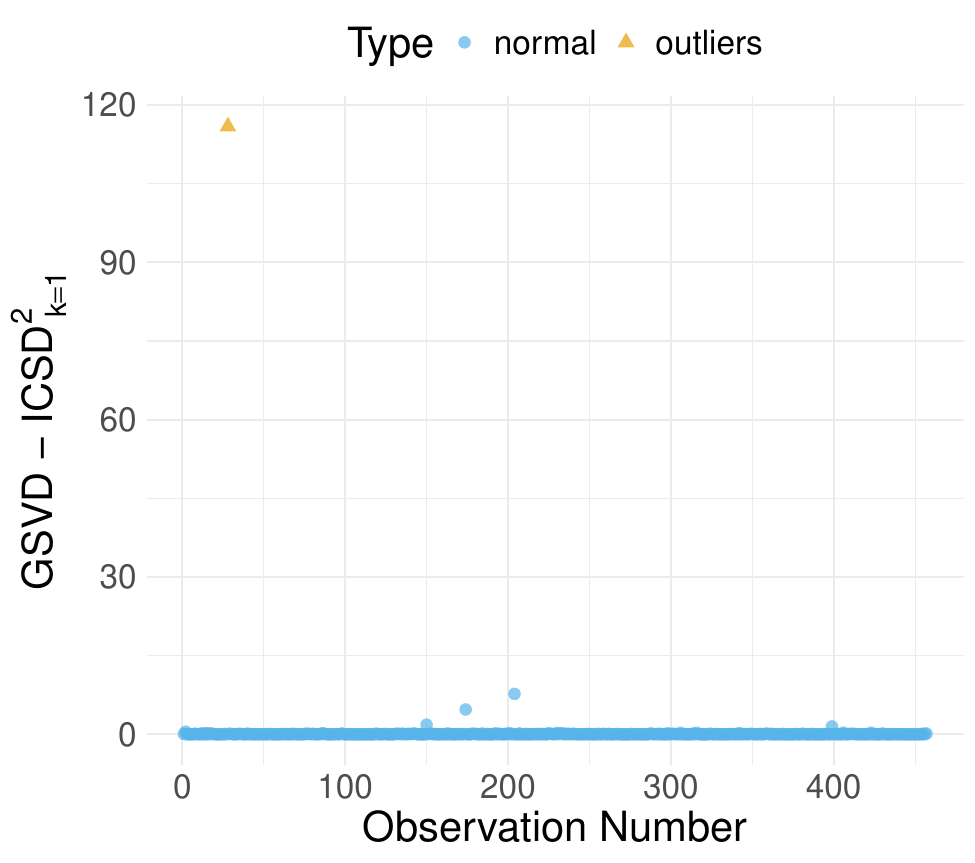}\hfill
			\includegraphics[keepaspectratio=true,width=5cm]{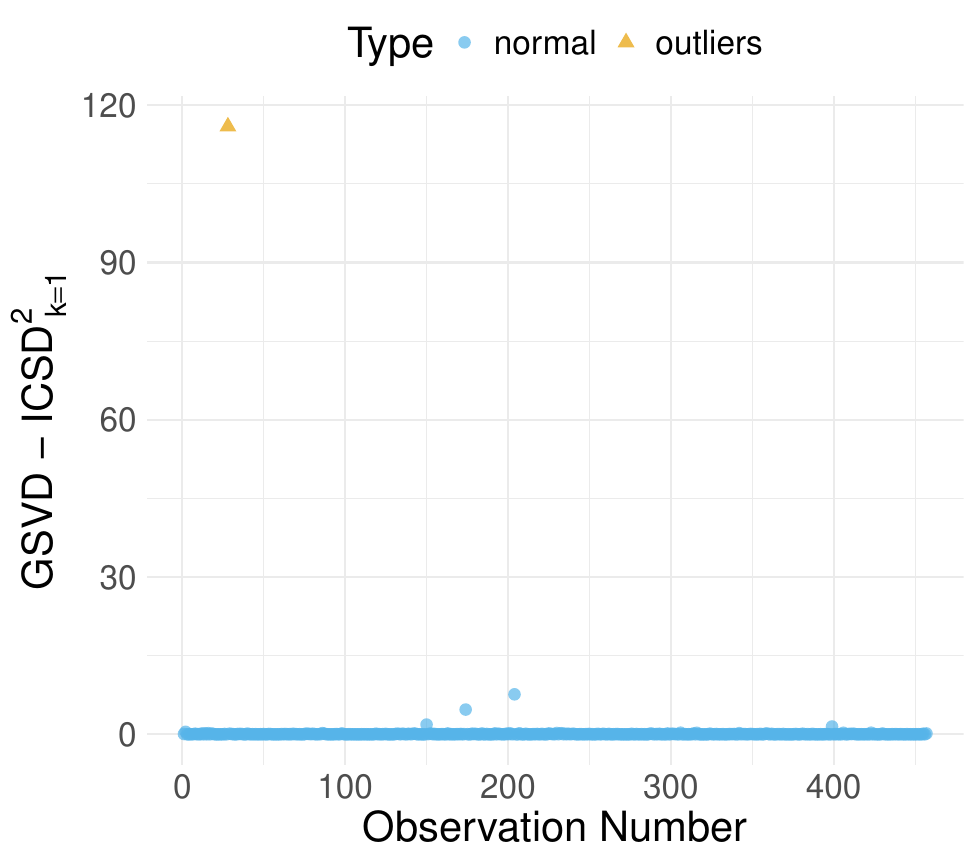}
			\caption{HTP2 data set: ICS distances, ICSD$^2$, computed with $k$ components for ICS of $\cov\cov_4$ after DR with $\rank(\X_n) = 138$ on initial data (1$^{st}$ panel), of $\cov-\covg_4$ with GSVD on initial data (2$^{nd}$ column) and on standardized data (3$^{rd}$ panel). The defective part is  in orange.}\label{fig:htp2}
\end{figure}

The rank estimation for a DR is quite challenging and unstable. It is estimated to 138 with the first criterion mentioned in Subsection~\ref{subsec:estim_rank}, 141 with the second and even to 1 with the third one based on the inertia. In addition, if the data is standardized then the two criteria lead to a rank of 141 and 51 for the third one. If we focus on the case of a rank of 138, then we cannot compute a robust scatter matrix like the $\mcd_{0.5}$. In Figure~\ref{fig:htp2}, we compute the ICSD$^2$ based on IC$_1$ and $\cov-\cov_4$ and the defective part is weirdly detectable as the observation having the smallest distance instead of the highest. However, doing it on the standardized data with 141 dimensions allows identifying it (see Figure~\ref{fig:htp2_append} in Appendix~\ref{append:applications}). This behavior shows that the DR approach is sensible to the estimation of the rank and the standardization of the data. On the contrary, if we perform ICS of $\cov-\covg_4$ with GSVD on the initial data (2$^{nd}$ plot) or on the standardized one (3$^{rd}$ plot), then the outlier is revealed and its ICSD$^2$ is stable between the two cases. It is noteworthy that the GSVD estimates 141 non-trivial eigenvalues. Finally, the regularized approach based on the $\mrcd$ does not identify the outlier (see Figure~\ref{fig:htp2_append} in Appendix~\ref{append:applications}).

\subsection{High Dimension Low Sample Size (HDLSS) case}\label{subsec:HDLSS}
\begin{figure}[h!]
			\includegraphics[keepaspectratio=true,width=5cm]{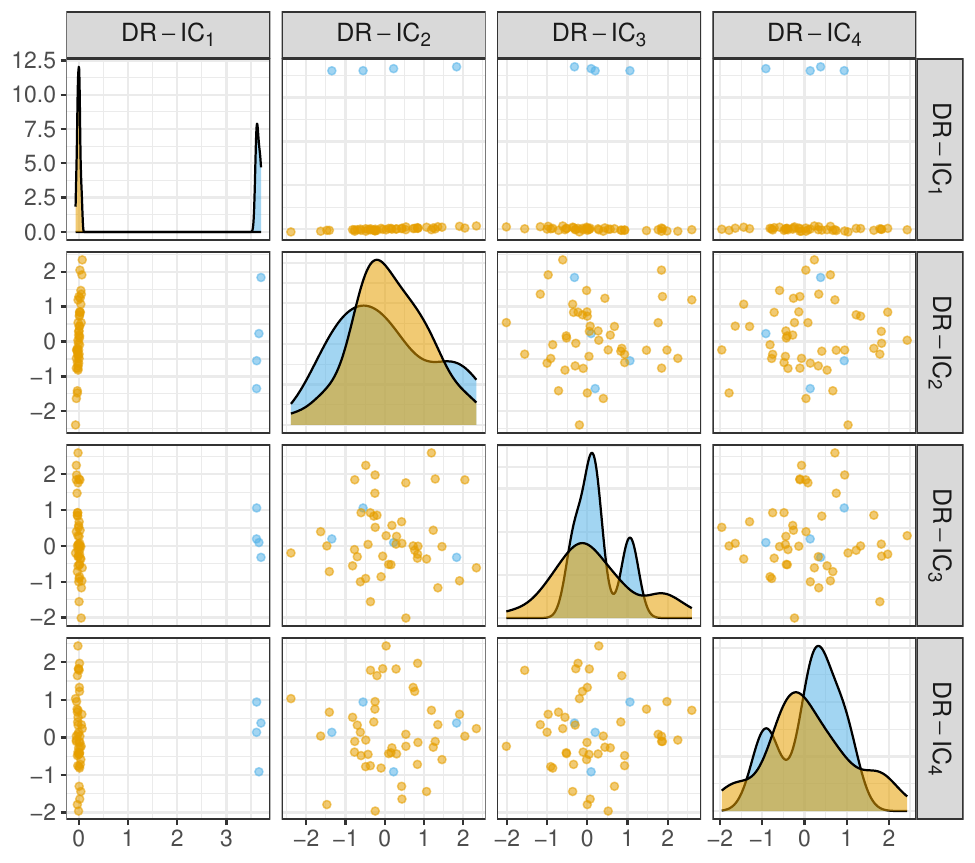} \hfill
			\includegraphics[keepaspectratio=true,width=5cm]{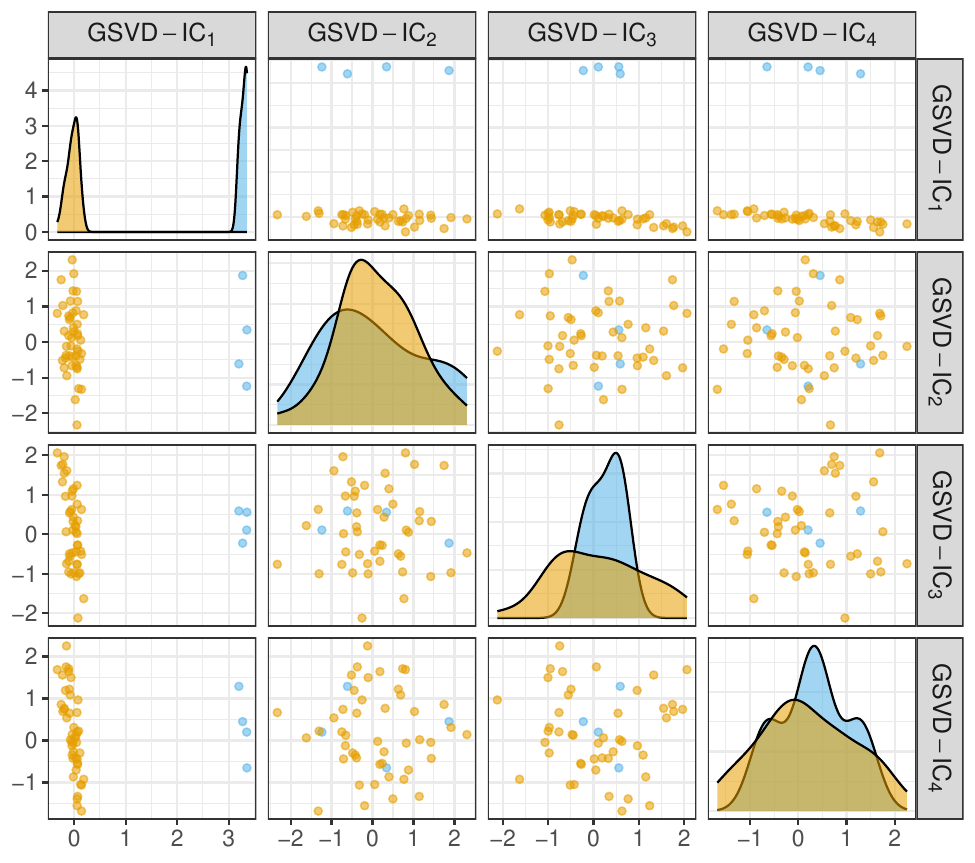}\hfill
   	\includegraphics[keepaspectratio=true,width=5cm]{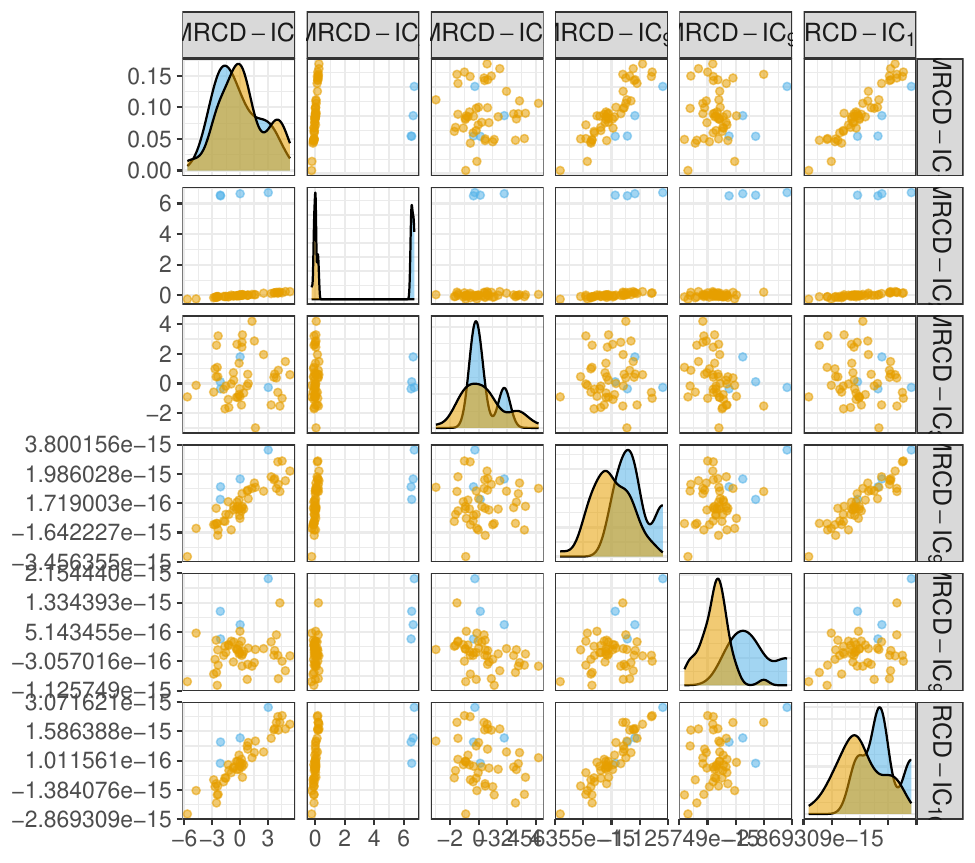}
				\caption{OC outliers in HDLSS: scatterplots matrix of the IC resulting of ICS of $\cov4-\cov$ after a dimension reduction with $\rank=4$ (on  1$^{st}$ column), of $\cov-\covg_4$ with GSVD (on 2$^{nd}$ column) and of $\mrcd-\cov$ (on 3$^{rd}$ column).}\label{fig:OC_HDLSS}
\end{figure}

To go further, we generate some data in an HDLSS context with more variables $p$ than observations $n$ but not lying in general position. Indeed, as already noted by 
\cite{lindsay2012fisher,bickel2018projection,pires2019high}, unless sparsity is enforced, projection pursuit or other statistical methods may detect an apparent structure that has no statistical significance. Following~\cite{she_robust_2016}, we generate $n=50$ observations on $p=100$ variables such as in Subsection~\ref{subsec:OC}. In this context, we retrieve similar results as when $n>p$: the classical ICS and GINV are returning an error, the results depend on the estimated rank and GSVD is working fine as we can see in Figure~\ref{fig:OC_HDLSS} for $\cov-\cov_4$ and $\cov-\covg_4$. For the robust scatter matrix $\mcd_{0.5}$ it is not possible to compute it on reduced data of 4 components and the outliers are not visible if we kept only two dimensions (see Figure~\ref{fig:OC_HDLSS_append} in Appendix~\ref{append:applications}). The last plot shows that $\mrcd-\cov$ identifies the outliers, but on IC$_2$ which is not what is expected.

\section{Conclusion\label{sec:conclu}}
Following previous ideas mainly used in  LDA context, we proposed three new ways of generalizing ICS to the case of positive semi-definite scatter matrices based on generalized inverse, dimension reduction or GSVD. We also investigated their theoretical properties (summarized in Table~\ref{tab:all}) and provided implementations in \proglang{R}. 
Theoretically, the approach based on GSVD looks the most appealing by keeping all the nice properties of the classical ICS. In practice, this method can only deal with a specific type of scatter matrices that has to be expressed as crossproducts and being affine equivariant. Relaxing this affine invariance property of ICS, empirical results showed interesting and stable results for different situations for GSVD with $\cov-\covg_4$, even on standardized data. For DR, estimating the rank of the data appears to be the main challenge with quite some sensibility on the results. In addition, the method is only orthogonal invariant and does not ensure that some robust scatter matrices can be computed on the reduced data. Finally, GINV seems the least suitable as it requires that $V_2$ can be computed on the initial data.

Overall, those methods allow to generalize ICS to the context of positive semi-definite scatter matrices and might be also helpful in the HDLSS case as long as the data are not in general position. Depending on the approach, it is important to think about which scatter should be the first one and if some OC outliers are present. In practice, it might be useful to perform ICS multiple times, exchanging $\V_1$ and $\V_2$ and trying the different implementations to compare the results or performing localized projection pursuit after ICS as suggested by \cite{dumbgen_refining_2023}. In the future, another idea could be to penalize or regularize ICS, as \cite{gaynanova_penalized_2017} or \cite{witten_penalized_2011} do for LDA for example.

\section*{Computational details}
All computations are performed with \proglang{R} version 4.3.3 \citep{r_core_team_r_2023} and uses the \textsf{R} packages \pkg{ICS}\citep{nordhausen_ics_2023} for ICS, \pkg{ICSClust}\citep{archimbaud_icsclust_2023}, \pkg{ICSOutlier}\citep{nordhausen_icsoutlier_2023}, \pkg{rrcov}\citep{todorov_rrcov_2024} for the MCD scatter matrix and \pkg{geigen}\citep{hasselman_geigen_2019} for computing GSVD.
Replication files are available from \url{https://github.com/AuroreAA/ICS_PSD_Replication}.

\section*{Acknowledgements}

This work is a generalization of the research conducted during my PhD under the supervision of Professor Anne Ruiz-Gazen whom I deeply thank for her guidance and insightful remarks on this topic. I also thank the associate editor and the two reviewers for their constructive and helpful remarks which greatly helped in increasing its quality.

\section*{Appendix A. \, Calculation details of the Moore-Penrose pseudo-inverse for Subsection~\ref{subsec:MP}}
 $\V(\X_n^*)^+$  has to satisfy the four conditions to be a Moore-Penrose pseudo-inverse:
	\begin{itemize}[noitemsep,topsep=0pt,parsep=0pt,partopsep=0pt]
		\item Condition 1: $ \V(\X_n^*) \V(\X_n^*) ^+ \V(\X_n^*)  = \V(\X_n^*)  $.		
		\item Condition 2: $\V(\X_n^*)^+\V(\X_n^*) \V(\X_n^*)^+ = \V(\X_n^*)^+ $.
		\item Condition 3: $(\V(\X_n^*) \V(\X_n^*)^+)^\top = \V(\X_n^*) \V(\X_n^*)^+$.
		\item Condition 4: $(\V(\X_n^*)^+ \V(\X_n^*))^\top = \V(\X_n^*)^+ \V(\X_n^*)$. 
	\end{itemize}
The proof of conditions 1 and 2 can be generalized to any matrix $\bo A$ but conditions 3 and 4 rely on the assumption of orthogonality of $\bo A$. 
Indeed, for condition 3, we have $(\V(\X_n^*) \V(\X_n^*)^+)^\top = (\bo A^\top)^{-1}\V_1(\X_n) \V_1(\X_n)^+ \bo A^\top$. Since $\bo A$ is orthogonal $(\bo A^\top)^{-1} = \bo A$ and $\V_1(\X_n)^+ = \bo A\V_1(\X_n)^+ \bo A^\top $,  so we obtained the desired equality:
$$(\bo A^\top)^{-1}\V_1(\X_n) \V_1(\X_n)^+ \bo A^\top= \bo A \V_1(\X_n) \bo A^\top \bo A \V_1(\X_n)^+ \bo A^\top= \V(\X_n^*) \V(\X_n^*)^+.$$
 The proof is similar for the condition 4.

\section*{Appendix B. \, Affine invariance for ICS with a Generalized Singular Value Decomposition for Subsection~\ref{subsec:GSVD}}\label{append:GSVD_aff}

\begin{proof}[\textbf{\upshape Proof:}] 
\textit{(i) Adaptation of the proof from \cite{tyler_invariant_2009}, appendix A.1, for distinct roots.}\\
Let $\X_n^* = \X_n \bo A +\bo 1_n \bo \gamma^\top$, with $\bs \gamma \in \mathcal{R}^p$. Then $\V_1(\X_n^*) = \bo A^\top \V_1(\X_n) \bo A$ and $\V_2(\X_n^*) = \bo A^\top \V_2(\X_n) \bo A$.\\
By definition of ICS we have, for $i = 1,\dots,p$:\\
 \begin{equation*}
    \begin{aligned}
					\alpha_i^2(\X_n^*) \V_2(\X_n^*) \bo b_i(\X_n^*)  &= \beta_i^2(\X_n^*) \V_1(\X_n^*) \bo b_i(\X_n^*),   \\
					\alpha_i^2(\X_n^*)  \bo A^\top \V_2(\X_n)  \bo A \bo b_i(\X_n^*)  &= \beta_i^2(\X_n^*)  \bo A^\top \V_1(\X_n)  \bo A \bo b_i(\X_n^*).\\
\text{Multiplying by $(\bo A^\top)^{-1}$:~~} \alpha_i^2(\X_n^*)  \V_2(\X_n)  \bo A \bo b_i(\X_n^*)  &= \beta_i^2(\X_n^*)  \V_1(\X_n)  \bo A \bo b_i(\X_n^*).
		\end{aligned} 
 \end{equation*}
If $\alpha_i^2(\X_n)/\beta_i^2(\X_n)$ is a distinct root, then $\alpha_i^2(\X_n)/\beta_i^2(\X_n) = \alpha_i^2(\X_n^*)/\beta_i^2(\X_n^*)$ and $\bo b_i(\X_n) \propto \bo A \bo b_i(\X_n^*)$, so $ \bo b_i(\X_n^*) \propto \bo A^{-1}  \bo b_i(\X_n)$ :
					$$\alpha_i^2(\X_n)  \V_2(\X_n)  \bo b_i(\X_n) = \beta_i^2(\X_n)  \V_1(\X_n)  \bo b_i(\X_n).$$
Projection onto $\bo b_i(\X_n^*)$:
					$\bo z_i^* = \bo b_i(\X_n^*)^\top \bo x^* = (\bo A^{-1}  \bo b_i(\X_n))^\top \bo A^\top \bo x = \bo b_i(\X_n)^\top \bo x = \bo z_i$.

\noindent{\textit{(ii) Proof from \cite{tyler_invariant_2009}, appendix A.1, for multiple roots.}}\\
In case of a multiple root of multiplicity $p_l$, the eigenvectors are not uniquely defined and can be chosen as any linearly independent vectors spanning the corresponding $p_l$-dimensional eigenspace. However the roots are still the same and so the subspace spanned by the corresponding $p_l$-dimensional eigenspace is still the same. 
\end{proof}

The case of multiple roots may appear  when $\V_1 \in  \mathcal{SP}_p$ and/or $\V_2 \in  \mathcal{SP}_p$ as it means than $\Null(\V_1) \neq \{0\}$ and/or $\Null(\V_2) \neq \{0\}$. For example, if we only focus on the cases where $\bo B \in \Null(\V_2) -   \Null(\V_1)$ or  $\bo B \in \Null(\V_1) -   \Null(\V_2)$, if $\dim(\Null(\V_2) -   \Null(\V_1)) >1$ and/or $\dim(\Null(\V_1) -   \Null(\V_2)) >1$ then $0$ and/or $\infty$ are multiple roots.

\section*{Appendix C. \, Additionnal results for the empirical applications in Section~\ref{sec:applications}}\label{append:applications}
\begin{figure}[h!]
			\includegraphics[keepaspectratio=true,width=5cm]{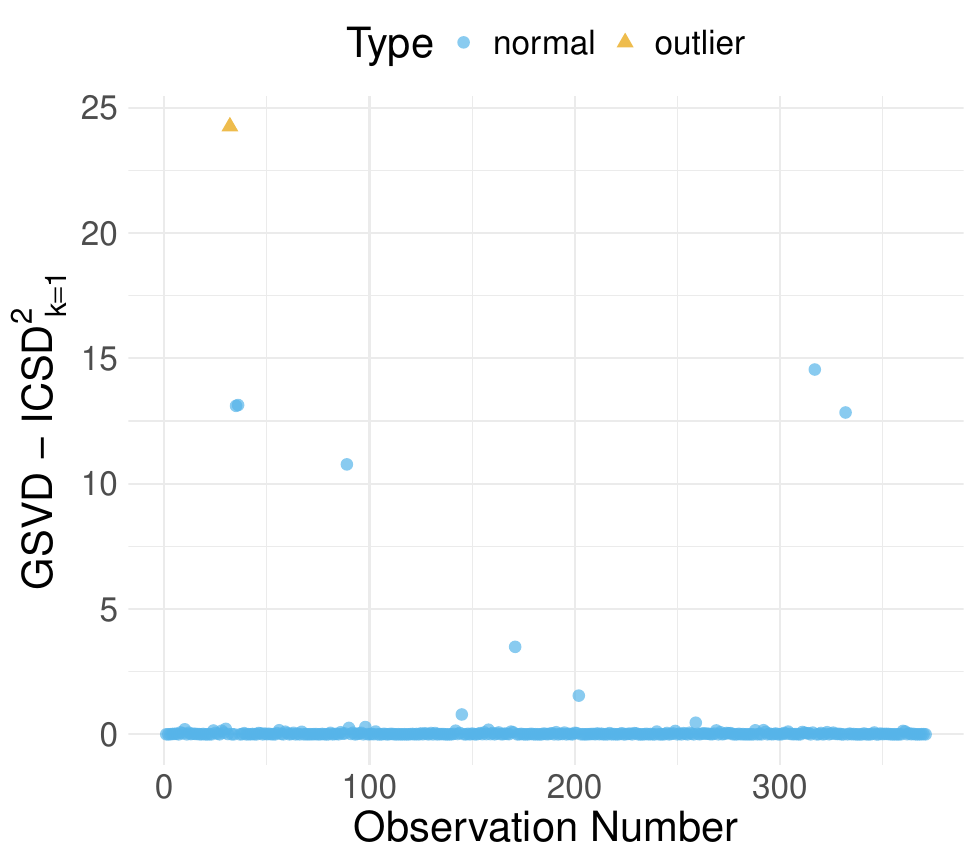} \hfill
			\includegraphics[keepaspectratio=true,width=5cm]{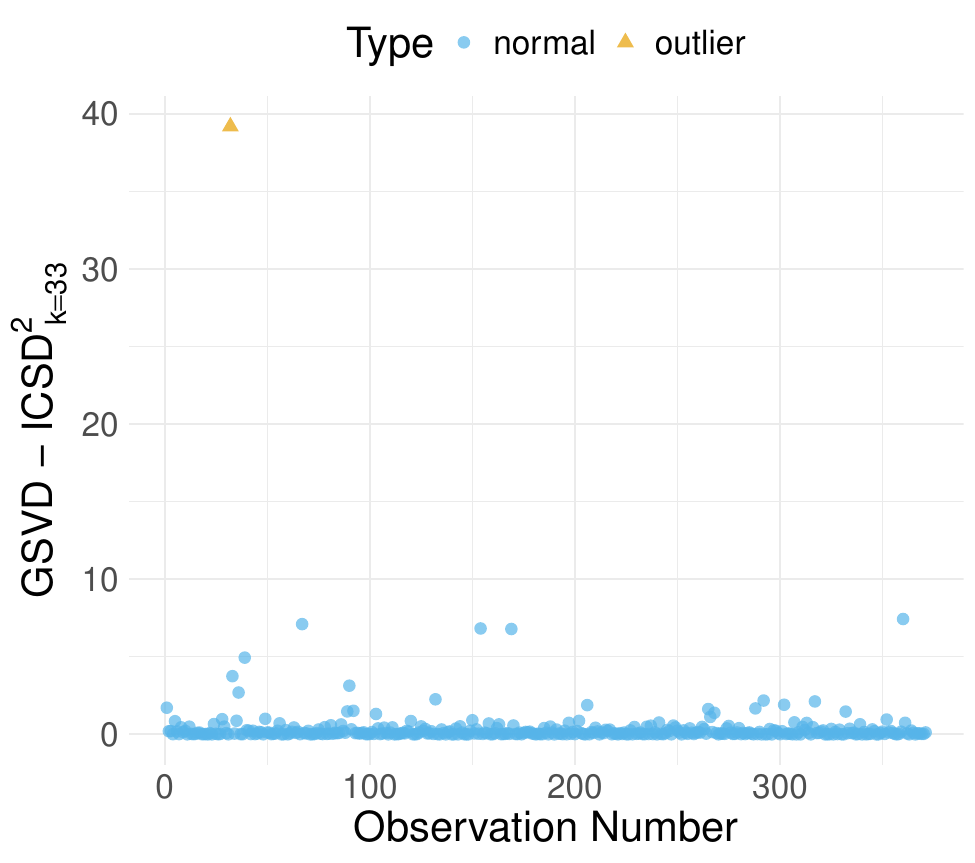}\hfill
			\includegraphics[keepaspectratio=true,width=5cm]{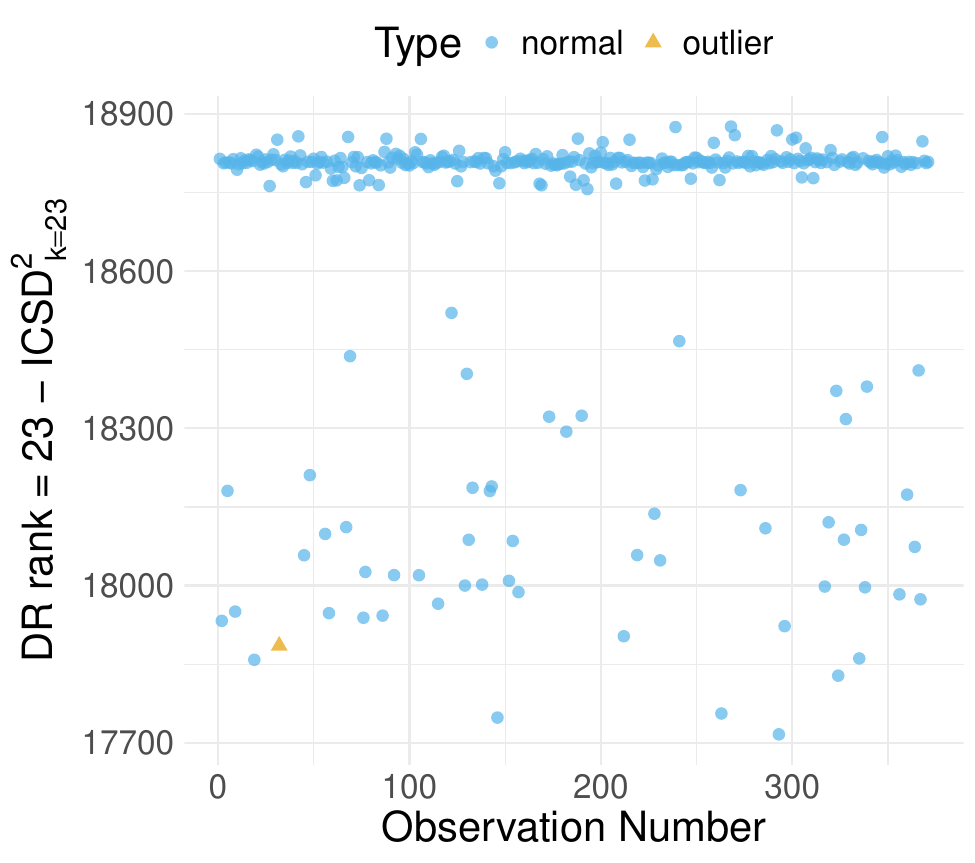}
		
			\caption{HTP3 dataset: ICS distances, ICSD$^2$, computed with $k$ components for ICS with GSVD of $\cov-\covg_4$ (1$^{st}$ panel), $\covg_4-\cov$  (2$^{nd}$ panel) and with $\cov-\mcd_{0.5}$ after DR (3$^{rd}$ panel). The defective part is  in orange.}\label{fig:htp3_append}
\end{figure}

\begin{figure}[h!]

			\includegraphics[keepaspectratio=true,width=5cm]{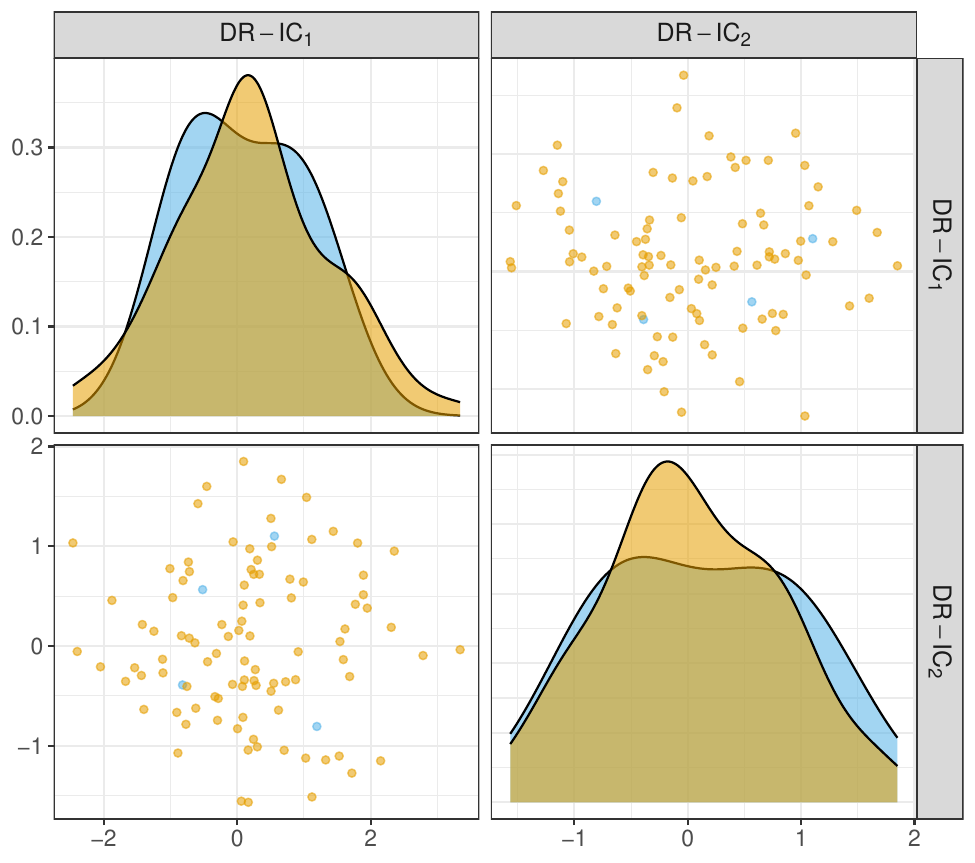}\hfill
			\includegraphics[keepaspectratio=true,width=5cm]{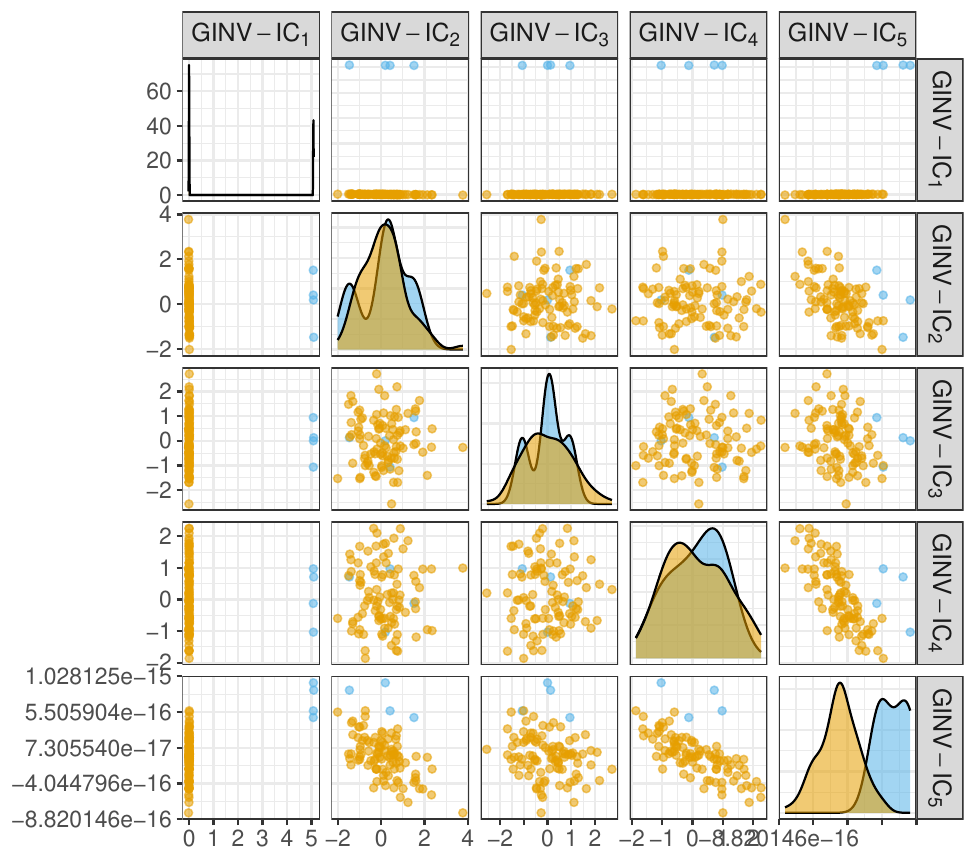} \hfill
			\includegraphics[keepaspectratio=true,width=5cm]{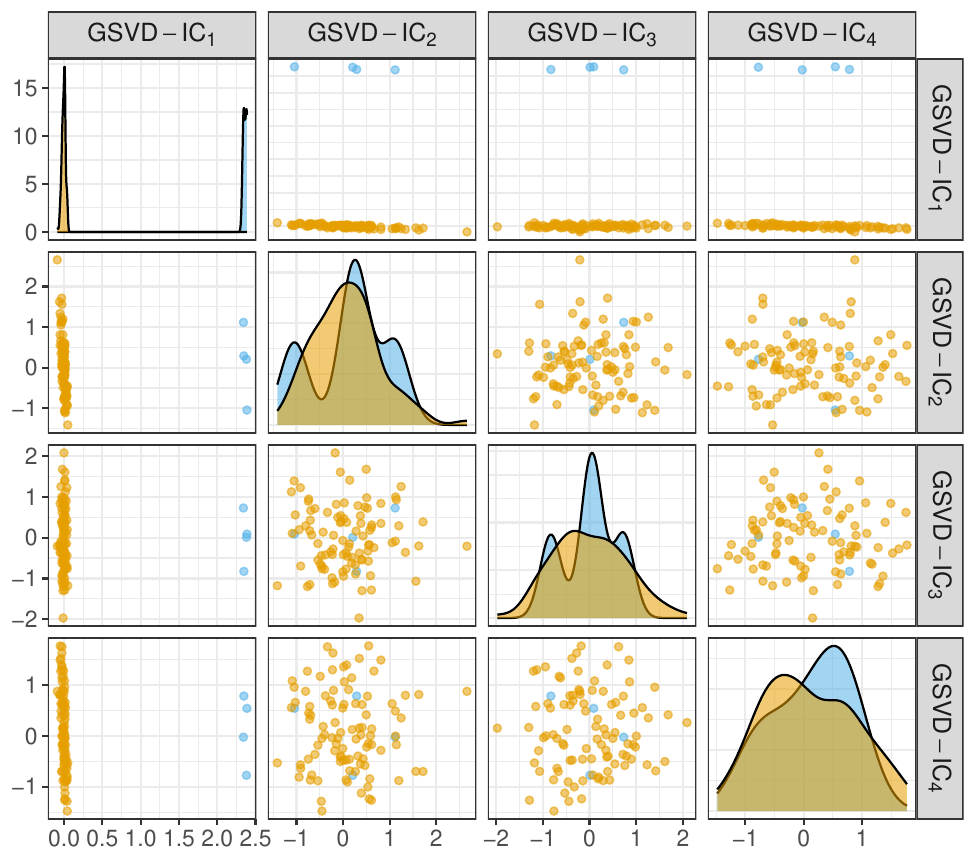}	
			\caption{OC outliers: scatterplots matrix of the IC resulting of ICS of $\mcd_{0.5}-\cov$ after a dimension reduction with $\rank=2$ (on  1$^{st}$ column), of $\cov-\cov_4$ with GINV (on 2$^{nd}$ column) and  of $\cov-\covg_4$ with GSVD (on 3$^{rd}$ column).}\label{fig:oc_append}
\end{figure}

\begin{figure}[h!]
			\includegraphics[keepaspectratio=true,width=5cm]{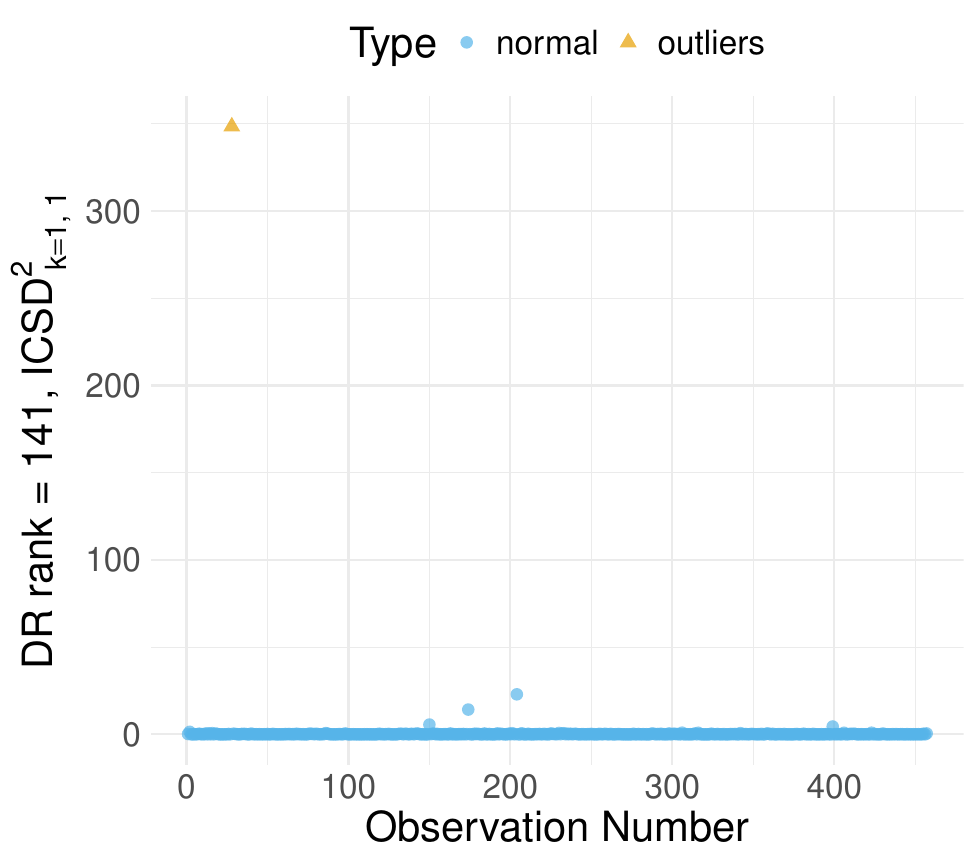} \hfill
			\includegraphics[keepaspectratio=true,width=5cm]{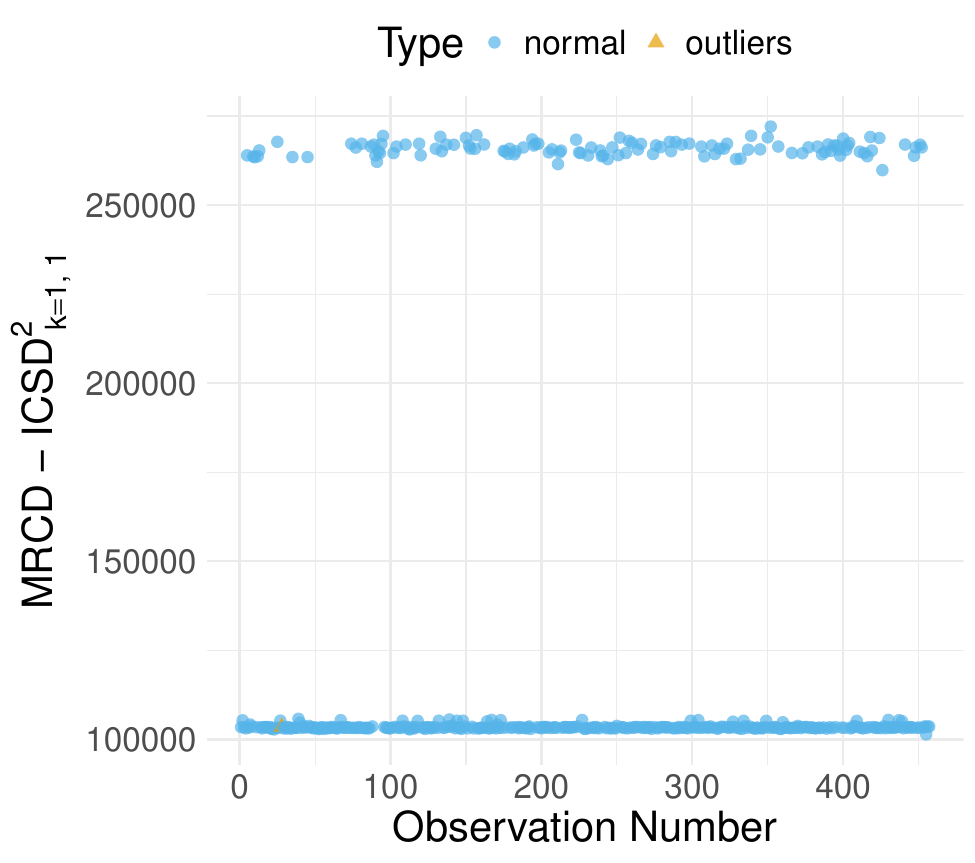}\hfill
			\includegraphics[keepaspectratio=true,width=5cm]{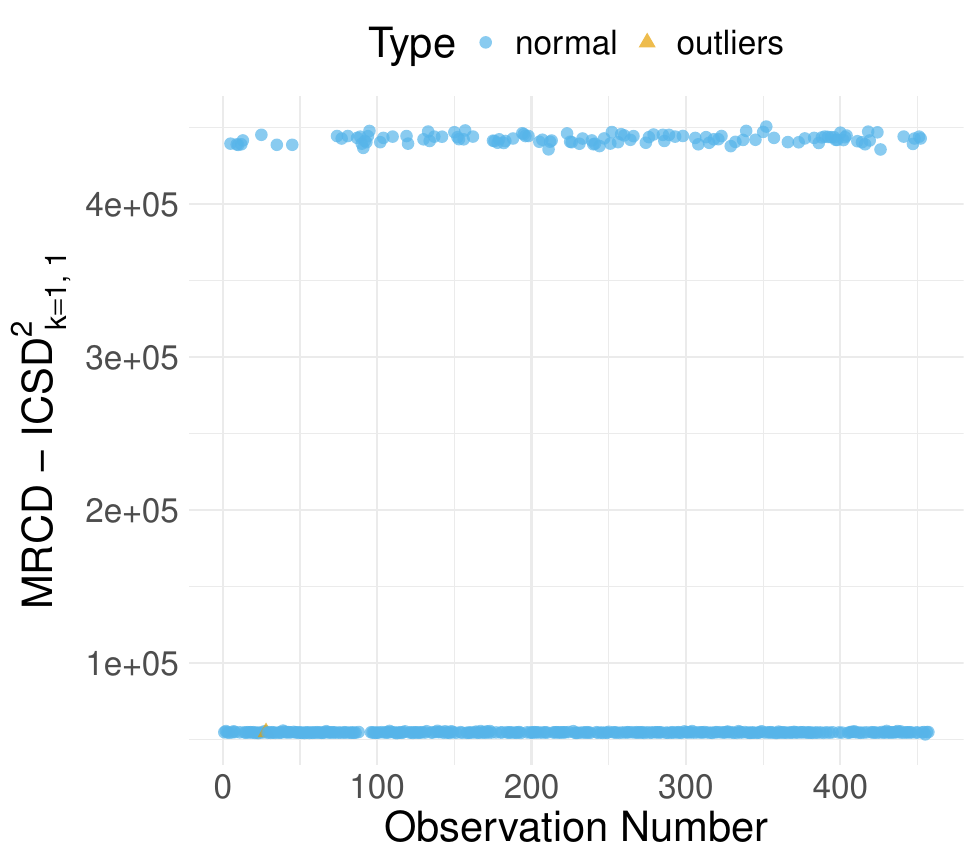}
			\caption{HTP2 data set: ICS distances, ICSD$^2$, computed with $k$ components for ICS of $\cov\cov_4$ after DR with $\rank(\X_n) = 141$ on standardized data (1$^{st}$ panel), of $\mrcd_{0.5}-\cov$ (2$^{nd}$ panel) and of $\mrcd_{0.5}-\cov$ on standardized data (3$^{rd}$ panel). The defective part is  in orange.}\label{fig:htp2_append}
\end{figure}

\begin{figure}[h!]
	\includegraphics[keepaspectratio=true,width=5cm]{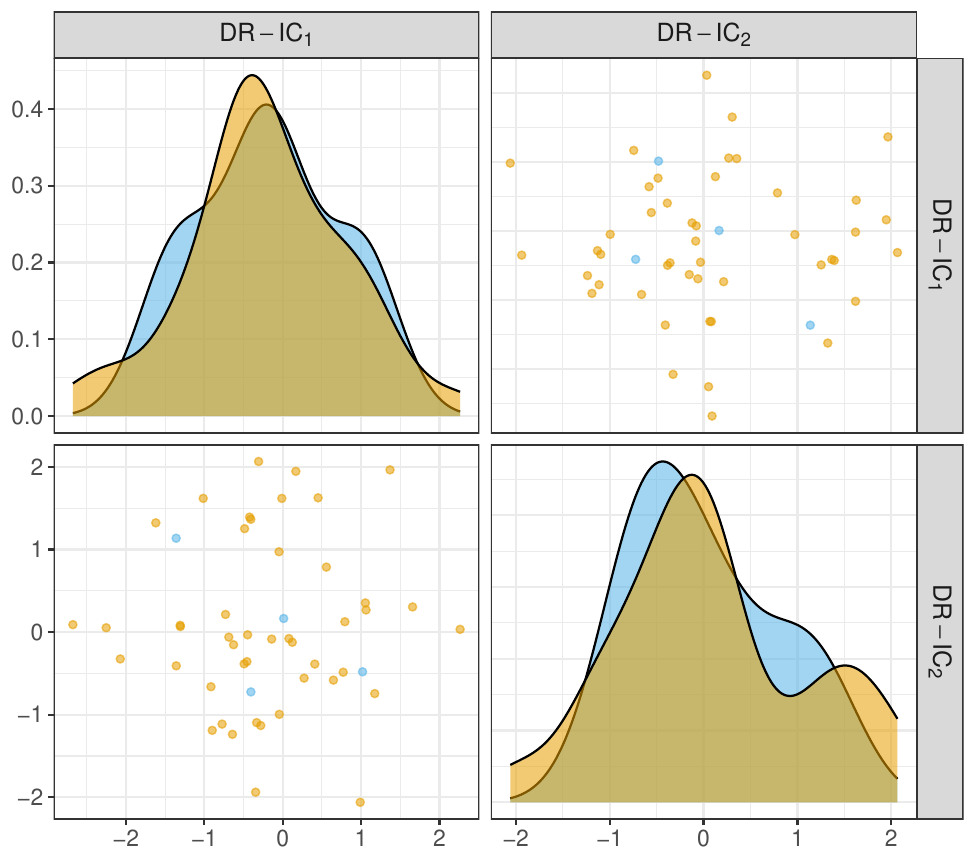} \hfill
			\includegraphics[keepaspectratio=true,width=5cm]{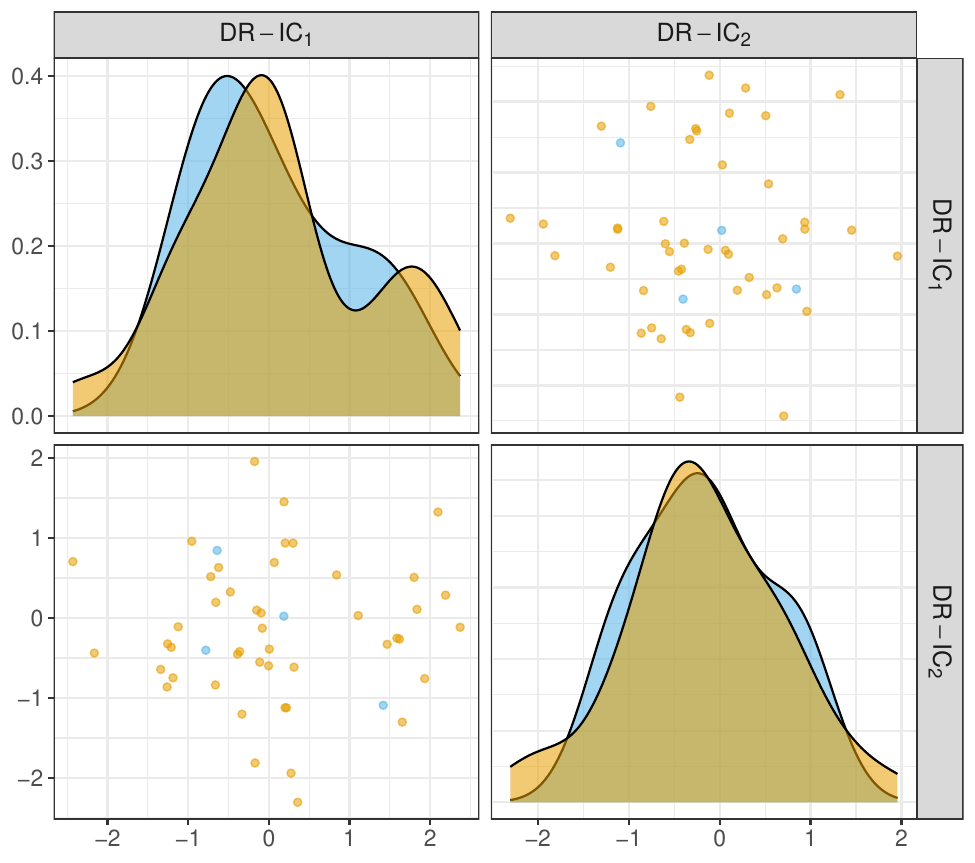} \hfill
			\includegraphics[keepaspectratio=true,width=5cm]{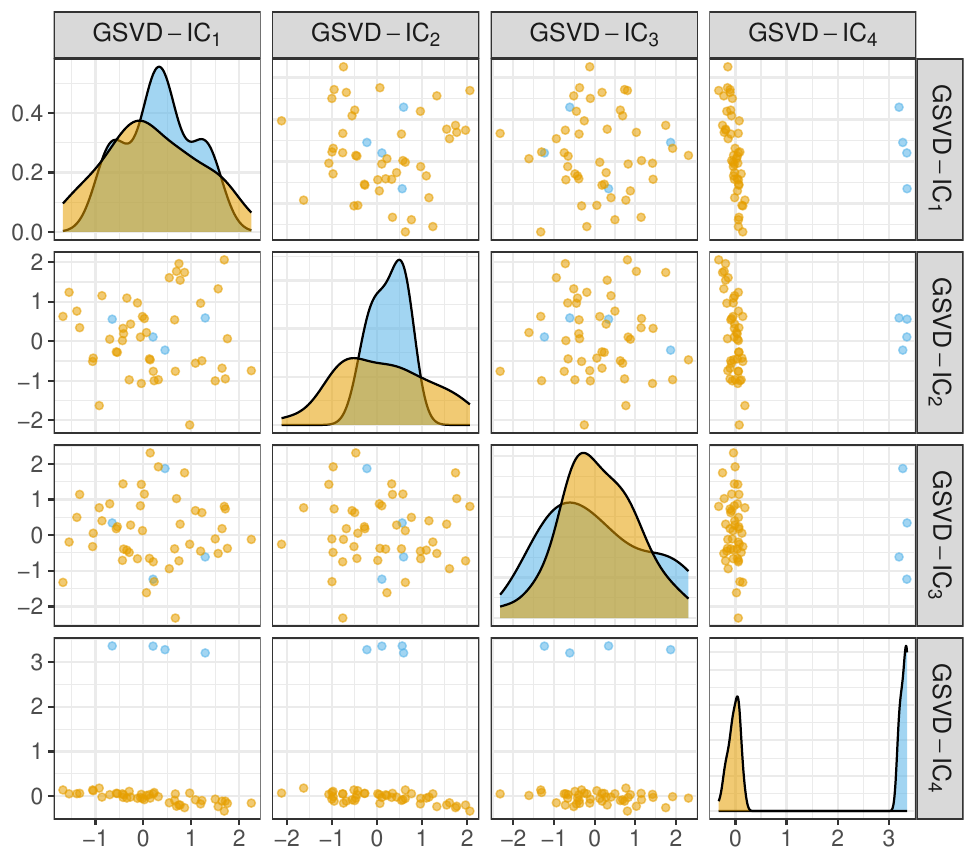}
				\caption{OC outliers in HDLSS: scatterplots matrix of the IC resulting of ICS of $\cov-\cov_4$ after a dimension reduction with $\rank=2$ (on  1$^{st}$ column), of  $\mcd_{0.5}-\cov$ (on 2$^{nd}$ column) and of $\covg_4-\cov$ with GSVD (on 3$^{rd}$ column).}\label{fig:OC_HDLSS_append}
\end{figure}

We display additional results regarding Subsection~\ref{subsec:nearly_col}
in Figure~\ref{fig:htp3_append}, Subsection~\ref{subsec:OC} in Figure~\ref{fig:oc_append}, 
Subsection~\ref{subsec:HTP2} in Figure~\ref{fig:htp2_append} and Subsection~\ref{subsec:HDLSS} in Figure~\ref{fig:OC_HDLSS_append}.

\bibliographystyle{myjmva}
\bibliography{references}

\end{document}